\documentclass[pdflatex,sn-basic,iicol]{sn-jnl}
\usepackage{graphicx}%
\usepackage{multirow}%
\usepackage{amsmath,amssymb,amsfonts}%
\usepackage{amsthm}%
\usepackage{mathrsfs}%
\usepackage[title]{appendix}%
\usepackage{xcolor}%
\usepackage{textcomp}%
\usepackage{manyfoot}%
\usepackage{booktabs}%
\usepackage{algorithm}%
\usepackage{algorithmicx}%
\usepackage{algpseudocode}%
\usepackage{listings}%
\usepackage{aas_macros}

\theoremstyle{thmstyleone}%
\theoremstyle{thmstyletwo}%
\theoremstyle{thmstylethree}%

\begin{document}

\title[]{Rapid quasi-periodic reconfiguration of the accretion column in pulsar 1A 0535+262}

\author*[1,3]{\fnm{Lingda} \sur{Kong}}\email{lingda.kong@nankai.edu.cn}

\author*[2,3]{\fnm{Xiaohang} \sur{Dai}}
\email{daixiaohang@ntsc.ac.cn}

\author[3]{\fnm{Andrea} \sur{Santangelo}}
\author[4]{\fnm{Long} \sur{Ji}}
\author[3]{\fnm{Valery F.} \sur{Suleimanov}}
\author[5,6]{\fnm{Alexander A.} \sur{Mushtukov}}
\author[3]{\fnm{Lorenzo} \sur{Ducci}}
\author[7]{\fnm{Shu} \sur{Zhang}}
\author[7,8]{\fnm{Qingcang} \sur{Shui}}
\author[7,8]{\fnm{Shuang-Nan} \sur{Zhang}}
\author[7]{\fnm{Hua} \sur{Feng}}
\author[9,3,7]{\fnm{Sergey S.} \sur{Tsygankov}}
\author[3]{\fnm{Honghui} \sur{Liu}}
\author[10,11]{\fnm{Pengju} \sur{Wang}}
\author[12]{\fnm{Qi} \sur{Liu}}

\affil[1]{\orgdiv{School of Physics}, \orgname{Nankai University}, \city{Tianjin}, \postcode{300071}, \country{China}}

\affil[2]{\orgdiv{National Time Service Center}, \orgname{Chinese Academy of Sciences}, \city{Xi'an}, \postcode{710600}, \country{China}}

\affil[3]{\orgdiv{Institut f{\"u}r Astronomie und Astrophysik, Kepler Center for Astro and Particle Physics}, \orgname{Eberhard Karls, Universit{\"a}t}, \orgaddress{\street{Sand 1}, \city{T{\"u}bingen}, \postcode{72076}, \country{Germany}}}

\affil[4]{School of Physics and Astronomy, Sun Yat-Sen University, Zhuhai, 519082, China}

\affil[5]{Mullard Space Science Laboratory, University College London, Holmbury St. Mary, Surrey RH5 6NT, UK}

\affil[6]{Astrophysics, Department of Physics, University of Oxford, Denys Wilkinson Building, Keble Road, Oxford OX1 3RH, UK}

\affil[7]{State Key Laboratory of Particle Astrophysics, Institute of High Energy Physics, Chinese Academy of Sciences, Beijing 100049, China}

\affil[8]{University of Chinese Academy of Sciences, Chinese Academy of Sciences, 100049, Beijing, China}

\affil[9]{Department of Physics and Astronomy, FI-20014 University of Turku, Finland}

\affil[10]{Research Center of Astronomy, QingHai University, Xining, 810016, PR China}

\affil[11]{Department of Physics and Astronomy, QingHai University, Xining, 810016, PR China}

\affil[12]{College of Physics, Hebei Normal University, Shijiazhuang 050024, China.}

\abstract{
Accretion onto strongly magnetized neutron stars is commonly interpreted using quasi-steady models, in which the accretion-column structure adjusts smoothly to the mass inflow rate. The cyclotron line in the X-ray spectrum, whose centroid energy traces the magnetic field strength and thus the height of the line-forming region, provides a key diagnostic of this structure. Whether this simple quasi-steady description remains valid on short dynamical timescales has remained uncertain.
Here we show that, during a giant outburst of the X-ray pulsar 1A~0535+262, quasi-periodic hard X-ray flux variations are accompanied by synchronized oscillations of the cyclotron line energy, with amplitudes exceeding those expected from simple accretion-rate fluctuations. 
The anti-correlation between cyclotron energy and apparent flux provides direct spectral-timing evidence for rapid changes in the line-forming region, which we interpret as geometric reconfiguration of the accretion column. The variability emerges in the luminosity regime where radiation pressure becomes dynamically important. These results reveal limitations of a simple quasi-steady interpretation for this source and suggest that radiation-supported columns can enter intrinsically dynamical states in high-luminosity accreting pulsars.}




\maketitle
\section{Introduction}\label{sec1}

Strongly magnetized accreting pulsars serve as unique astrophysical laboratories for probing the dynamics of matter and radiative transfer under extreme gravitational and magnetic fields. Despite decades of multi-method observational and theoretical efforts, fundamental mechanisms governing accretion physics in X-ray pulsars (XRPs) remain actively debated \citep{Mushtukov2022arXiv220414185M}. In particular, key unresolved questions concern the dynamical structure of accretion columns and the physical processes regulating radiation production and transport within them \citep{Becker1998ApJ...498..790B, Mushtukov2015MNRAS.447.1847M, Becker2022ApJ...939...67B}, as well as their coupling to the surrounding magnetosphere.

Hydrostatic models of accretion columns predict distinct deceleration mechanisms and radiation beam patterns across different accretion regimes \citep{Davidson1973NPhS..246....1D, Basko1976MNRAS.175..395B, Burnard1991ApJ...367..575B, Becker2012AA...544A.123B}. These theoretical expectations have gained broad observational support through luminosity-dependent evolution of the X-ray continuum \citep{Reig2013AA...551A...1R, Kong2020ApJ...902...18K}, pulse-phase-resolved spectroscopy \citep{Wang2022ApJ...935..125W, Xiao2024ApJ...963...42X}, cyclotron resonance scattering feature (CRSF) behaviour \citep{Staubert2019AA...622A..61S, Kong2021ApJ...917L..38K}, and, more recently, polarized emission variability \citep{Doroshenko2022NatAs...6.1433D}. Within this framework, the centroid energy of the CRSF is widely interpreted as a tracer of the magnetic field strength and the characteristic height of the line-forming region in the accretion column, such that long-term CRSF--luminosity correlations reflect changes in column geometry driven by variations in the mass accretion rate \citep{Becker2012AA...544A.123B}. An implicit assumption underlying this widely adopted paradigm is that the accretion column responds quasi-steadily to changes in the mass accretion rate, so that its structure at any given time is uniquely determined by the instantaneous accretion rate. Under this assumption, variability observed on different timescales is expected to follow the same CRSF--luminosity relation. Whether this quasi-static description remains valid on short timescales, however, has not been directly tested observationally.

Low-frequency quasi-periodic oscillations (LFQPOs), commonly observed in strongly magnetized accreting pulsars, provide a natural opportunity to probe accretion dynamics on dynamical timescales. 
LFQPOs typically manifest as millihertz peaks in power density spectra and exhibit diverse phenomenology across different sources and accretion states. Their time duration, frequency evolution, amplitudes and energy dependence vary substantially in different sources (see details in supplementary materials), suggesting that multiple physical processes may contribute to their production. This has left the physical origin of LFQPOs unsettled, largely because a general, model-independent approach to systematically investigate their properties has been lacking.
Furthermore, the characteristic timescales of LFQPOs are orders of magnitude shorter than those associated with global luminosity evolution, making them a probe of rapid, non-stationary processes in the inner accretion flow and the accretion column itself. 

Because the CRSF directly traces the magnetic-field strength and emission geometry in the accretion region, while the LFQPO probes short-timescale dynamical variability, a natural approach is to examine how the CRSF evolves over the QPO cycle.
By combining phase-resolved spectroscopy with non-stationary signal analysis techniques (VMD-HHT method) that recover the instantaneous frequency and phase of the QPO, this strategy directly links timing variability to physical changes in the accretion column itself. We suggest that such a joint timing–spectral framework provides a broadly applicable method for investigating the origin of LFQPOs in systems that exhibit cyclotron lines, and more generally for probing non-stationary accretion under extreme magnetic fields.

Among known accreting X-ray pulsars, the Be/X-ray binary 1A~0535+262 represents a particularly well-suited laboratory for this purpose. The system ($d \approx 2$ kpc; \citealt{Bailer-Jones2018}) hosts a highly magnetized neutron star with a spin period of $P_{\rm spin}=104$~s, orbiting an O9.7IIIe companion in an eccentric ($e=0.47\pm0.02$) and wide ($P_{\rm orb}=110.3\pm0.3$~days) orbit \citep{Steele1998, Finger1996ApJ...459..288F}. The presence of CRSFs at $\sim45$~keV and $\sim100$~keV indicates a surface magnetic field strength of $\sim5\times10^{12}$~G \citep{Kong2021ApJ...917L..38K}. This source exhibits persistent LFQPOs during giant outbursts, with pronounced energy dependence at hard X-ray energies \citep{Finger1996ApJ...459..288F, Ma2022MNRAS.517.1988M}, as well as well-characterized long-term CRSF--luminosity correlations \citep{Kong2021ApJ...917L..38K, Shui2024MNRAS.528.7320S}. These combined properties make 1A~0535+262 an exceptional target for investigating non-stationary accretion processes and the internal dynamics of accretion columns.

In this work, we analyze Insight–HXMT observations of the 2020 giant outburst of 1A~0535+262 and perform QPO phase–resolved spectroscopy to probe the accretion column on dynamical timescales. 
We report a quasi-periodic modulation of the CRSF centroid energy synchronized with the QPO cycle, with a behaviour that is qualitatively distinct from the established long-term CRSF–luminosity relation. The CRSF energy varies coherently within individual QPO cycles, indicating that the line-forming region changes on timescales of tens of seconds and in a manner decoupled from the global accretion-rate evolution. These results provide direct spectral-timing evidence for rapid changes in the line-forming region, which we interpret as geometric reconfiguration of the accretion column, and reveal limitations of a simple quasi-steady description for this source on QPO timescales.

\section{Results}\label{sec3}

\subsection{QPO Frequency Evolution and Energy Dependence}

In order to study the QPO properties and to perform further QPO phase-resolved spectral analysis in the following sections, we reanalyzed the Insight–HXMT observations of the 2020 outburst of 1A~0535+262.
To improve the statistical quality of the timing analysis, observations obtained within the same day were combined. Quasi-periodic oscillation (QPO) searches were then systematically performed on all available observations from November 6 to December 24. 

The power density spectrum (PDS) is dominated by a coherent neutron-star spin signal at $\sim104$~s and its harmonics, producing a series of sharp narrow peaks. To prevent contamination of the QPO search in the 30--90~mHz band, the average pulse profile was removed from the original light curve before computing the PDS. The spin period was identified using the $Z_n^2$ search method provided by \texttt{Stingray} \citep{Buccheri1983AA...128..245B}. 
The PDSs were computed from background- and pulse-subtracted HE light curves in the 25–80keV band and the 1/512–0.5Hz range, normalized to units of (rms)$^{2}$~Hz$^{-1}$, corrected for the expected white-noise level, and rebinned using a geometrical series.
The PDSs can be modeled well with three Lorentzian components: two broad components describing the low- and high-frequency noise and one narrow component for the QPO. The observations with detected QPOs ($>3\sigma$) are listed in Supplementary Table 1 and plotted in Fig.~\ref{QPO_fit}.

Observations of the 2020 outburst of 1A~0535+262 with Insight-HXMT (MJD 59159--59207) revealed quasi-periodic oscillations (QPOs) above 25~keV throughout the outburst, except below $\sim 2 \times 10^{37}$~erg~s$^{-1}$ (Fig.~\ref{QPO_fit}), and the QPO frequency increased with flux during the outburst. Below such luminosity, no significant QPOs were detected. We evaluated the significance of a possible QPO component around 20--30~mHz using an F-test. The resulting F-statistic values were generally low ($\ll 3\sigma$), indicating that a QPO component could not be distinguished from the underlying noise. Compared with the giant outburst in 1994 monitored by BATSE, the peak luminosity of the 2020 giant outburst was much higher. The maximum QPO frequency increased from 72~mHz to 94~mHz. More intriguingly, in the 1994 giant outburst, we also found that the QPO disappeared at the onset or the end of the outburst when the flux was lower than $\sim 1 \times 10^{-8}$~erg~cm$^{-2}$~s$^{-1}$ in 20--100~keV (Fig.~13 and Table~2 in \cite{Finger1996ApJ...459..288F}). We therefore conclude that the inferred $\nu_{\rm QPO}$--$L_X$ relation is not affected by incomplete sampling or observational bias.

In Fig.~\ref{QPO_fit}, the QPO frequency scales with the observation-averaged X-ray luminosity as $\nu_{\rm QPO}\propto L_{37}^{0.62\pm0.01}$ (red line), indicating a link to the secular evolution of the accretion flow during the outburst. This luminosity is averaged over each observation and is used as a proxy for the long-term mean mass-accretion rate, not for the apparent flux variations within an individual QPO cycle. The mHz QPOs are often interpreted within the framework of the Keplerian Frequency Model (KFM; \citealt{Klis1997ASSL..218..121V}) or the Beat Frequency Model (BFM; \citealt{Alpar1985Natur.316..239A}). In the KFM, QPOs arise from periodic obscuration by inhomogeneities at the inner disk edge, such that $\nu_{\rm QPO}=\nu_{\rm K}(r_{\rm in})$. In the BFM, blobs orbiting at the Keplerian frequency at the disk–magnetosphere boundary are accreted at a rate modulated by the neutron-star spin, producing a beat frequency $\nu_{\rm QPO}=\nu_{\rm K}(r_{\rm in})-\nu_{\rm s}$. Under the canonical dipolar magnetospheric scaling with a luminosity-independent disk--magnetosphere coupling factor, both models predict a much shallower dependence, close to $\nu_{\rm QPO}\propto L_{37}^{3/7}$, and provide poor fits to the measured frequency–luminosity relation (see details in the supplementary material). We therefore disfavor the canonical fixed-coupling KFM/BFM implementations, while noting that more generalized disk--magnetosphere frequency models require additional assumptions and cannot be excluded by the slope alone. The QPO fractional rms was calculated in seven energy bands at three different luminosity levels, which are shown in Fig.~\ref{sub_fitting_rms}d--f. All QPOs show similar and strong energy dependence with a hump near the CRSF energy. Our results are consistent with the results in \cite{Ma2022MNRAS.517.1988M}.

The combined evidence of a steep frequency–luminosity scaling, a hard X-ray–confined RMS peak near the CRSF energy, and the QPO-phase CRSF modulation is difficult to reproduce with canonical disk-boundary interpretations alone. Instead of being explained solely by accretion-rate modulation or obscuration at the disk–magnetosphere interface, the QPO likely reflects intrinsic radiative and geometric variability within the accretion column itself. In the following sections, we explore this possibility and its implications for the observed spectral–timing behaviour.

\subsection{QPO Phase Resolved Spectral Analysis}

We further investigate the coupling between the QPO phase and spectral variability to constrain the physical origin of its unusual energy and luminosity dependence. Using the Hilbert–Huang transform based on Variational Mode Decomposition (VMD-HHT), the instantaneous phase (IP) of a quasi-periodic oscillation (QPO) signal can be extracted and applied to phase folding, resulting in a QPO profile. Following this, we divided the QPO cycle into 10 equal phases, extracted the energy spectrum for each phase, and conducted a spectral analysis. Notably, since the distribution of pulsar phases is approximately uniform (see Supplementary Fig. 13) in each QPO phase, the results of the QPO phase-resolved spectral analysis are not affected by pulsar phase modulations.

The spectral fitting was performed over 2--10~keV (LE), 10--30~keV (ME), and 30--150~keV (HE), with the continuum described by a power law modified by a Fermi–Dirac cutoff (\texttt{FDcut}) \citep{Tanaka1986LNP...255..198T}: $I_E = K \cdot E^{-\Gamma} \left[ 1 + \exp\left( \frac{E - E_{\text{cutoff}}}{E_{\text{fold}}} \right) \right]^{-1}$, combined with the Tuebingen–Boulder interstellar medium absorption model (\texttt{TBabs}) \citep{Wilms2000ApJ...542..914W}. The hydrogen column density $n_{\rm H}$ is fixed at $0.44 \times 10^{22}$~cm$^{-2}$ based on the H$\mathrm{I}$ 4 Pi Survey \citep{HI4PI2016AA...594A.116H}. A constant is used to cross-calibrate the responses of different detectors. Additionally, we included a Gaussian emission line (\texttt{gaussian}) at approximately 6.7~keV and two Gaussian absorption structures (\texttt{gabs}) at $\sim 45$~keV and $\sim 100$~keV, representing the iron emission line and the cyclotron resonant scattering features (CRSFs), respectively. The fitting results for three observations with different flux levels using \texttt{constant$\times$TBabs$\times$gabs$\times$gabs$\times$(gaussian+FDcut)} for the average spectrum and two QPO phases at trough and peak are shown in Supplementary Tables 2-4. The parameters modulated with QPO phases are shown in Supplementary Figs 6-8.
In Fig.~\ref{sub_fitting_rms}, panels (a--c), the best spectral fits and residuals for the QPO phase 0.5--0.6 (blue points) and phase 0.0--0.1 (red points) for three observations are shown. The main spectral differences are clearly related to the 25--85~keV range, aligning with the absence of QPO signals below 25~keV. We found that the normalization of the model does not change across all three observations. In Obs.~1, the photon index $\Gamma$, cutoff energy $E_{\rm cut}$, and folding energy $E_{\rm fold}$ exhibit a positive correlation with the QPO profile. However, $\Gamma$ and $E_{\rm cut}$ do not show obvious variations in Obs.~2 and Obs.~3 (for Obs.~3, $E_{\rm cut}$ is fixed at zero), and only $E_{\rm fold}$ keeps the same trend.

The line energy $E_{\rm cyc1}$ and line depth $S_{\rm cyc1}$ of the fundamental CRSFs in all three observations exhibit anti-correlated sinusoidal modulation with the QPO flux. Notably, the line energy and width of the CRSFs do not exhibit any significant correlations or degeneracies with $\Gamma$, $E_{\rm cut}$, and $E_{\rm fold}$ of the continuum model. As an example, the parameter distributions from the spectral fitting of QPO phase 0.5--0.6, where the strength of the CRSF is weakest, on November 20, derived using Markov Chain Monte Carlo (MCMC) with a chain length of 100,000, are shown in Supplementary Fig. 12. Furthermore, considering that \texttt{cyclabs} can describe asymmetric CRSF shapes in some cases, we tested this model by replacing \texttt{gabs} while keeping the continuum unchanged. We find that the main results, including the QPO-phase modulation of the CRSF centroid energy, remain consistent within uncertainties (see the supplementary materials). Therefore, the modulation is not sensitive to the choice of the CRSF line profile, and the effect arising from the asymmetric CRSF shape is excluded.

To test the robustness of the CRSF modulation against the choice of continuum model, we performed spectral fits using both phenomenological and physically motivated continua. Specifically, we fitted both the phase-averaged and QPO phase–resolved spectra with the physical model \texttt{COMPMAG}, which describes emission from thermal and bulk Comptonization in a cylindrical accretion column above the polar cap of a magnetized neutron star \citep{Farinelli2012AA...538A..67F}. The overall continuum part of the spectra can be fitted reasonably well by \texttt{COMPMAG} and \texttt{bbodyrad}. However, the number of free parameters in this model significantly exceeds what can be robustly constrained by the available data, particularly in the context of QPO phase–resolved spectroscopy. As a result, strong parameter degeneracies arise, preventing a unique determination of the physical parameters. To mitigate this issue, we restricted the seed photon blackbody temperature ($kT_{\rm bb}$), optical depth of the accretion column ($\tau$), the index of the velocity profile ($\eta$), the terminal velocity at the NS surface ($\beta_0$), and the radius of the accretion column ($r_0$) to physically reasonable ranges, and fixed them, except for the electron temperature ($kT_{\rm e}$) and normalization ($R_{\rm km}^2/D_{10}^2$), to the parameters of the phase-averaged spectrum in the QPO phase–resolved fits. Under these conditions, the modulation of the CRSF energy with QPO phase remains statistically significant. The results are shown in Supplementary Tables 5-7. The parameters modulated with QPO phases are shown in Supplementary Figs 9-11

To calculate the significance level of the sinusoidal modulation of the CRSF energy, we used a non-linear least-squares method to fit a sinusoidal model to the data over one period, accounting for measurement errors. The fitting process involved minimizing the chi-square statistic. Using an F-test, we compared the fits of the sinusoidal model and a constant model. Fig.~\ref{spectra_fit} indicates that the confidence level of this modulation at the fundamental CRSF central energy is approximately 5.1~$\sigma$ in Obs.~1, 3.9~$\sigma$ in Obs.~2, and 4.0~$\sigma$ in Obs.~3. The line energy $E_{\rm cyc2}$ of the harmonic CRSFs also shows an anti-correlation with the QPO flux, but the trend is not statistically significant ($<3\sigma$) due to the limited photon statistics above 100~keV (see the supplementary materials).

Using the spectral fitting model, we can estimate the root mean square (RMS) of the flux modulated by QPO phases across different energy bands, shown as black lines in Fig.~\ref{sub_fitting_rms}(d--f): 
$f_{\text{rms}} = \frac{\sqrt{\sum_{i=1}^{N} (F_i - \overline{F})^2/N}}{\overline{F}}$, 
where $N=10$ is the total number of QPO phase bins, $\overline{F}$ is the phase-averaged flux, and $F_i$ is the phase flux. This result closely matches the fractional rms derived by integrating the QPO component in the energy-dependent power density spectra (PDS) in the frequency domain. Furthermore, the blue line calculated from the fitting model excluding the CRSFs fails to reproduce the energy dependence of the fractional RMS. This clearly indicates that the CRSF modulation plays a crucial role in the broad peak structure of the RMS energy dependence. 

\begin{figure}[h]
    \centering
    \includegraphics[width=0.49\textwidth]{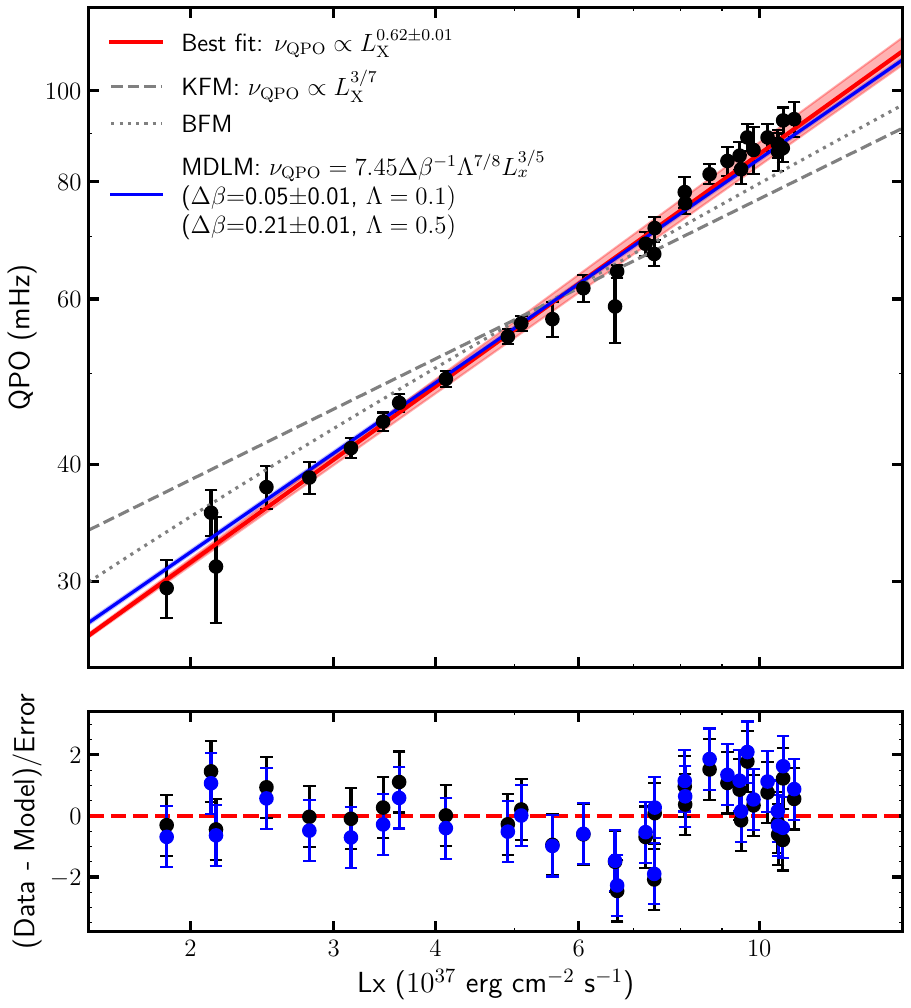}
    \caption{
    QPO Frequency Evolution with X-ray Luminosity. The black points in the top panel show the observed QPO frequency as a function of X-ray luminosity ($L_{\mathrm{X}}$). The solid red line with uncertainty band indicates the best-fit model $\nu_{\rm QPO}\propto L_{\rm X}^{0.62\pm0.01}$. The dashed and dotted lines represent the correlation predicted from the Keplerian Frequency Model (KFM) and Beat Frequency Model (BFM), respectively. The solid blue line with its uncertainty band shows the best fit of the magnetic field deformation and mass leakage model (MDLM) prediction to the full $\nu_{\rm QPO}$--$L_X$ dataset. The bottom panel displays the residuals for the best-fit model (black points) and the MDLM (blue points). The fitting results demonstrate that $\nu_{\rm QPO}$ increases more rapidly than predicted by the KFM and BFM, in agreement with the MDLM prediction.
    }
    \label{QPO_fit} 
\end{figure}

\begin{figure*}
    \centering
    \includegraphics[width=0.9\textwidth]
    {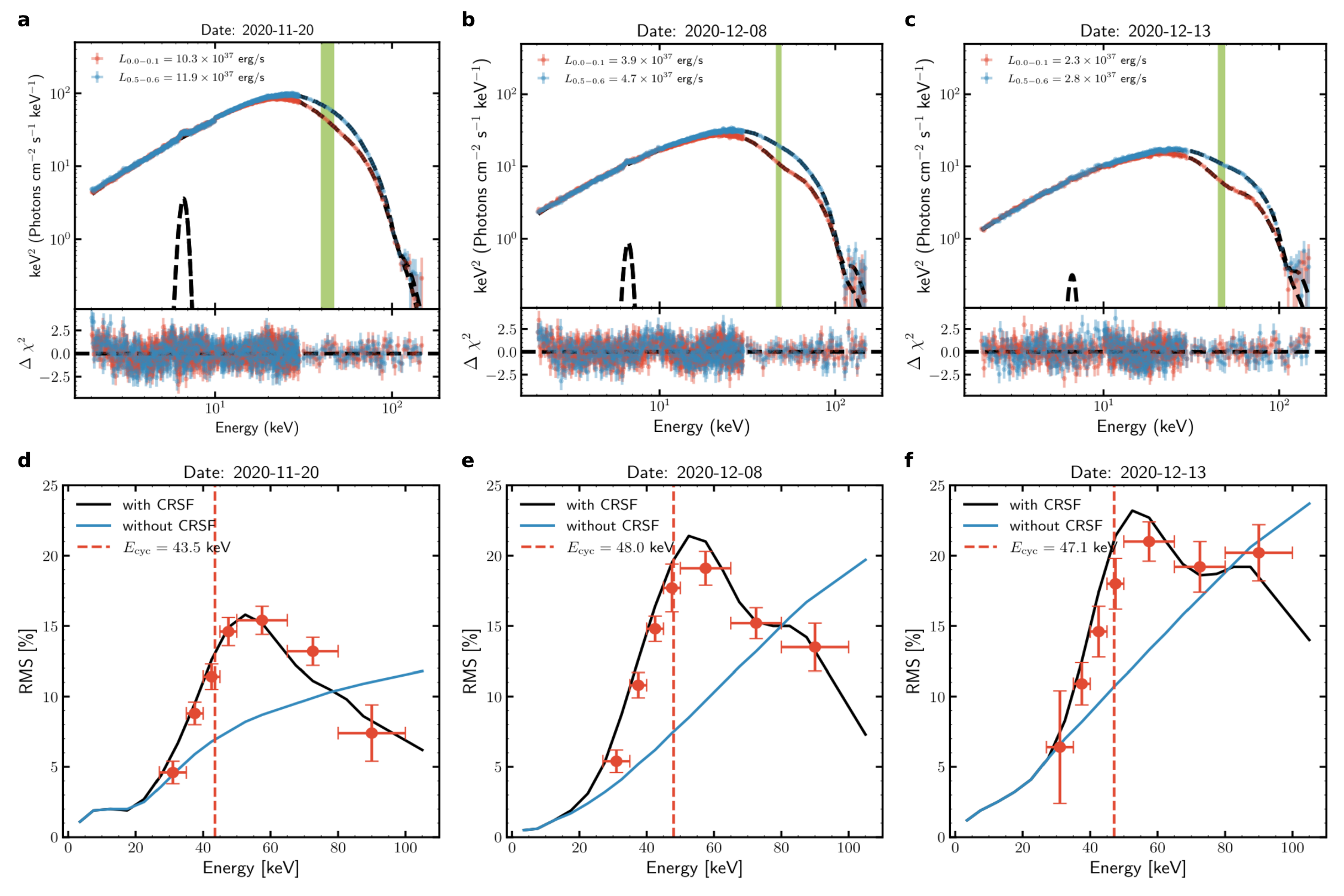}
    \caption{Spectral fitting and energy dependence of the QPO Fractional RMS from Observations on November 20, December 8, and December 13, 2020. (a--c) X-ray spectra of the source for the three observation dates, fitted with models including a cyclotron resonance scattering feature (CRSF). The spectra are shown in the $\nu^2 F_{\nu}$ representation, where the red and blue points indicate the observational data in different QPO phases. The residuals of the fits are displayed in the lower panels. (d--f) The fractional root mean square (RMS) of the QPO as a function of energy for the same three observations. The data points (red markers) are obtained directly from the power density spectrum (PDS) analysis. The black and blue curves represent the RMS derived from the flux of the models, with and without the inclusion of the CRSF component, respectively. The red dashed line indicates the centroid energy of the CRSF. These results suggest a strong correlation between the CRSF and the energy-dependent QPO variability.}
    \label{sub_fitting_rms} 
\end{figure*}

\begin{figure*}
    \centering
    \includegraphics[width=0.9\textwidth]{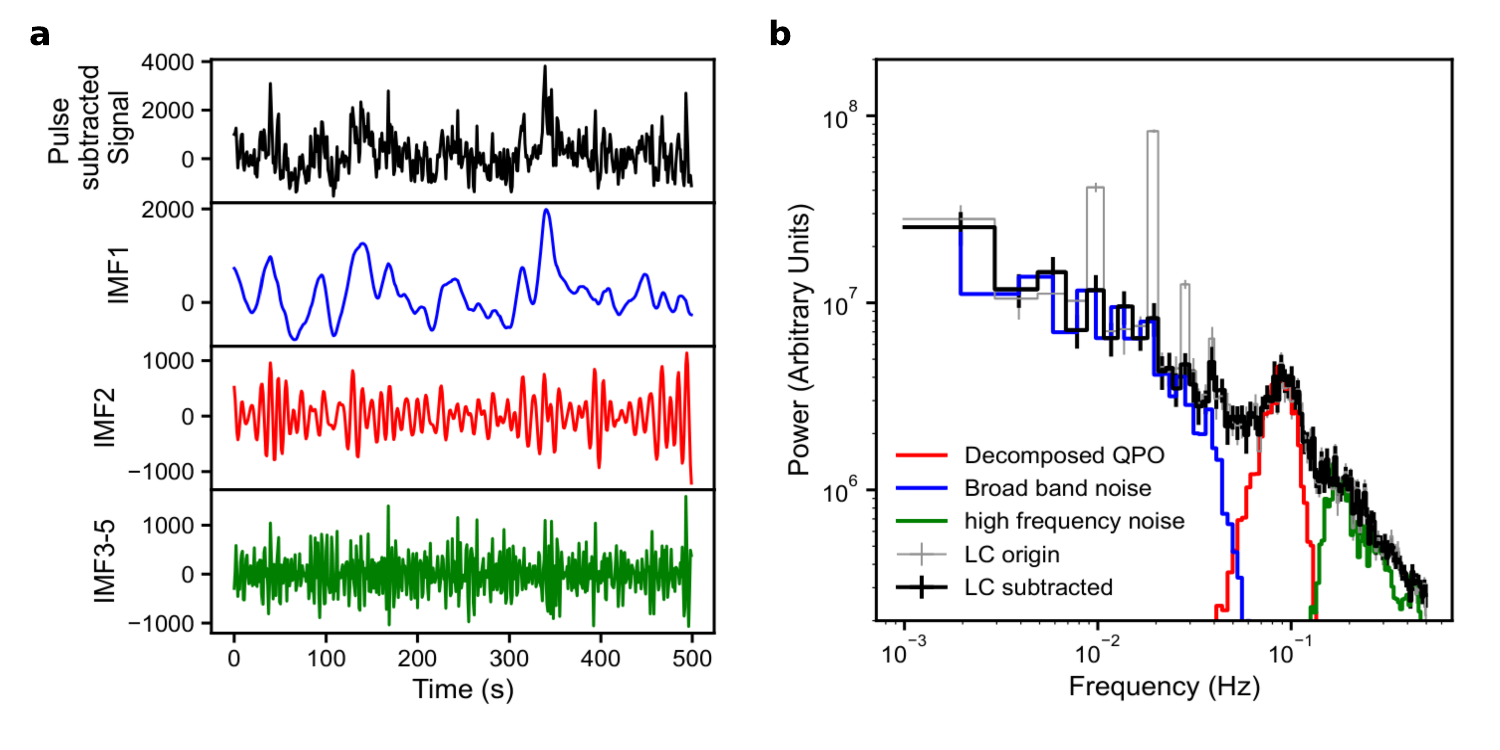}
    \caption{
    Decomposition of the light curve using the Variational Mode Decomposition (VMD) method. (a) The 25--80~keV pulse-subtracted light curve (black line) is decomposed into five intrinsic mode functions (IMFs). The extracted IMF1 (blue) does not contain phase-coherent pulse information. Instead, it represents stochastic accretion-rate fluctuations commonly observed in accreting X-ray pulsars and can be appropriately categorized as low-frequency broadband noise. IMF2 (red) corresponds to the quasi-periodic oscillation (QPO) signal. The summation from IMF3 to IMF5 (green) is associated with high-frequency noise. (b) Power density spectrum (PDS) analysis. The PDS of the pulse-subtracted light curve can be well described by the decomposed broadband noise (blue), the isolated QPO component (red), and the high-frequency noise component (green), which are derived from IMF1, IMF2, and IMF3--5, respectively. The gray line represents the PDS of the original light curve before pulse profile subtraction for comparison. Subtracting the pulse profile from the light curve does not affect the continuum shape of the PDS.
    }
    \label{PDS_fit} 
\end{figure*}

\begin{figure*}[h]
    \centering
    \includegraphics[width=0.9\textwidth]{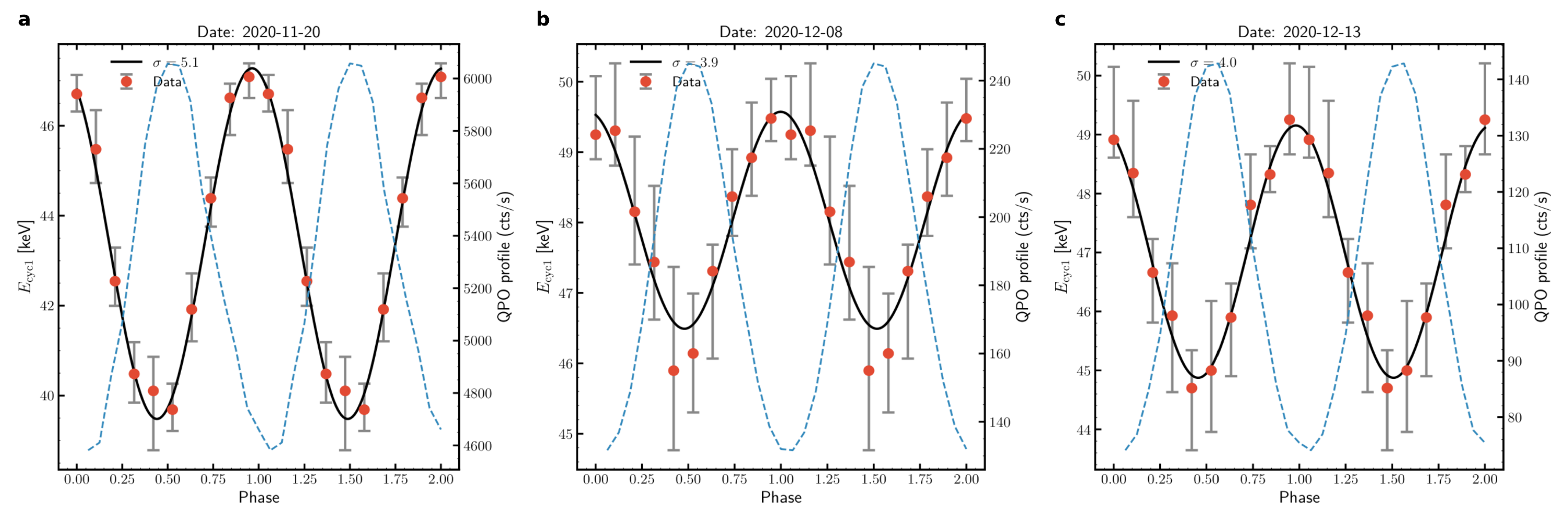}
    \caption{
    QPO-phase modulation of the fundamental CRSF energy during the 2020 outburst of 1A~0535+262. The fundamental cyclotron resonance scattering feature (CRSF) centroid energy, $E_{\mathrm{cyc}}$, is shown as a function of QPO phase for three representative observations: November~20 (a), December~8 (b), and December~13 (c). The red data points represent the CRSF energies obtained from QPO phase--resolved spectral fitting, with vertical error bars indicating $1\sigma$ uncertainties. The black solid curves show the best-fitting sinusoidal modulation models. The blue dashed curves denote the normalized QPO profiles, plotted for reference and scaled to the right-hand axis in each panel. A statistically significant sinusoidal modulation of $E_{\mathrm{cyc}}$ is detected in all three observations, with the CRSF energy anti-correlated with the QPO flux. The modulation significance reaches $5.1\sigma$ on November~20 and remains significant at $3.9\sigma$ and $4.0\sigma$ on December~8 and December~13, respectively. These observations reveal a statistically significant, phase-coherent modulation of the CRSF energy associated with the QPO cycle, suggesting rapid changes in the accretion-column emission region.
}
    \label{spectra_fit} 
\end{figure*}

\section{Discussion}\label{sec4}

The results above place several simultaneous constraints on the QPO mechanism in 1A~0535+262. The canonical KFM/BFM prescriptions based on a fixed dipolar magnetospheric scaling provide poor descriptions of the frequency–luminosity evolution (see the supplementary materials). In addition, the energy dependence of the QPO modulation strength (RMS) cannot be simply attributed to variations in the Comptonized continuum component. Instead, the hump of the RMS spectrum arises primarily from changes in the energy and depth of the cyclotron resonance scattering feature. The sinusoidal-like modulation pattern of the CRSF energy indicates rapid, coherent changes in the line-forming region. We interpret this behaviour as geometric variability within the accretion column, while emphasizing that the geometry is inferred indirectly through CRSF spectral-timing diagnostics rather than imaged directly.

\begin{figure*}
    \centering
    \includegraphics[width=0.9\textwidth]{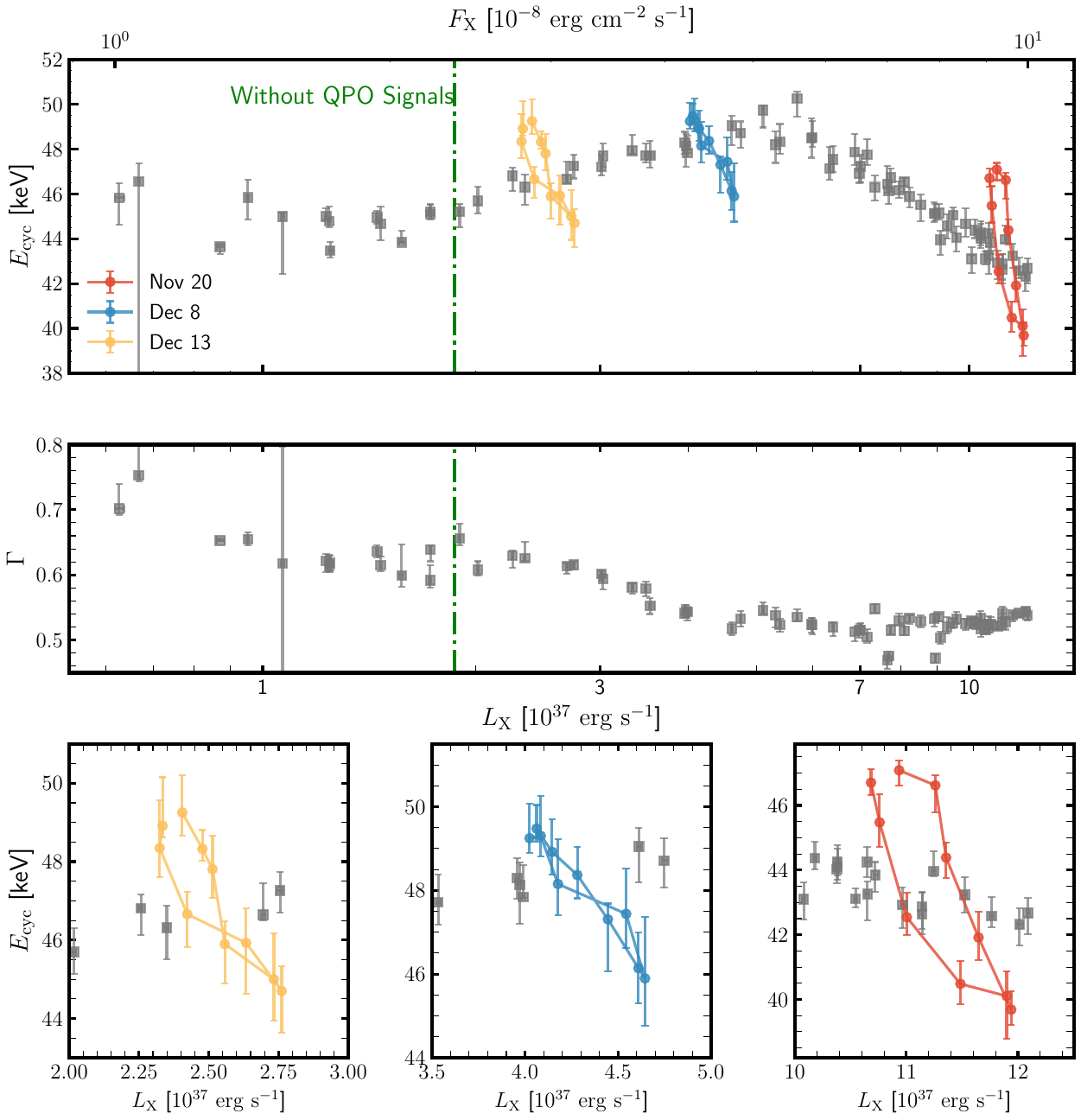}
    \caption{
Top panel: The long-term evolution of the CRSF centroid energy with luminosity during the 2020 giant outburst of 1A~0535+262. The colored points represent the modulation of the CRSF energy and the corresponding apparent luminosity across ten QPO phases observed on November~20 (red), December~8 (blue), and December~13 (orange), 2020. The vertical green dashed line marks the luminosity boundary below which QPO signals are absent. Note that the colored points should not be interpreted as a short-timescale extension of the long-term CRSF--luminosity relation. The CRSF modulation associated with the QPO phases exhibits a steeper slope and a different trend compared to the long-term CRSF--luminosity relation. These QPO-related luminosity variations primarily reflect changes in the measured luminosity induced by CRSF variations, rather than genuine short-timescale changes in the intrinsic mass accretion rate. Middle panel: The evolution of the photon index $\Gamma$ with the Fermi–Dirac cutoff power-law model (\texttt{fdcut}). 
After the QPO appears, $\Gamma$ decreases with increasing luminosity, indicating spectral hardening that is qualitatively consistent with the radiation-dominated column calculations of \citet{Becker2022ApJ...939...67B}. Bottom panels: Zoomed-in views of the epochs of the three QPO observations.
}
    \label{Long_Short}
\end{figure*}

We compare the CRSF centroid energy as a function of luminosity on both long and short timescales (Fig.~\ref{Long_Short}). The gray points trace the long-term pulse-to-pulse CRSF--luminosity evolution across the entire 2020 outburst, while the colored points show the CRSF variations across QPO phases for three representative observations. On long timescales, the CRSF energy evolves smoothly with luminosity, following the well-known bimodal behaviour observed in subcritical/supercritical accretion columns. Our results agree with previous works within the overlapping luminosity range \citep{Klochkov2011AA...532A.126K, Becker2012AA...544A.123B}, while the additional behaviour detected at higher luminosities extends the predicted bimodal CRSF--luminosity framework into a previously unexplored regime\citep{Kong2021ApJ...917L..38K}. 
On QPO timescales, however, the CRSF energy exhibits a much steeper, phase-coherent modulation, clearly deviating from the long-term trend. Here, we need to emphasize that the luminosity variation in Fig.~\ref{Long_Short} with the QPO phase may reflect different physical scenarios compared to the average luminosity value. The former might merely be a change in apparent flux calculated from spectral fitting model caused by variations in the CRSF component, and does not necessarily reflect the accretion rate variation during the QPO period within a short time scale. However, for the latter case, we still consider its possibility.
Under the conventional quasi-steady paradigm, the accretion column structure is uniquely determined by the instantaneous mass accretion rate, and the CRSF--luminosity relation should therefore be invariant across timescales. In the supplementary materials, we applied the supercritical column model of \citet{Becker2012AA...544A.123B} to estimate the emission-height variation induced solely by changes in $\dot{M}$. Even adopting reasonable values for the effective photon transport parameter, the predicted CRSF shifts are significantly smaller than the observed QPO-phase modulation. Reproducing the $\sim8$~keV variation would require emission-height changes far exceeding those allowed by hydrostatic adjustment to accretion-rate fluctuations alone. Some alternative mechanisms are considered as well. Relativistic Doppler effects predict a positive correlation between CRSF energy and flux, contrary to the observed anti-correlation. Reflection models can reproduce long-term negative CRSF--luminosity trends in some sources, but they cannot account for the large, phase-coherent modulation amplitude without invoking similarly extreme and rapid geometric changes. 

These considerations once again indicate that the short-timescale CRSF modulation cannot be interpreted as a quasi-steady luminosity-driven response. Instead, the luminosity variations associated with the QPO do not trace genuine changes in the mass accretion rate, but arise primarily from spectral redistribution linked to the cyclotron feature itself (see Fig.~\ref{sub_fitting_rms}). Interestingly, QPOs are detected only above a characteristic luminosity threshold. During both the 1994 and 2020 outbursts, they disappear at lower luminosities near the onset and decay phases. In Fig.~\ref{Long_Short}, this boundary coincides with the luminosity regime where the positive CRSF--luminosity relation begins, commonly interpreted as the onset of a radiation-dominated shock and radiation braking in the accretion column (see Fig.~1c in \cite{Becker2012AA...544A.123B}).
The middle panel of Fig.~\ref{Long_Short} further shows that, after the QPO appears, the photon index $\Gamma$ decreases as luminosity increases, indicating progressive spectral hardening. This behavior is qualitatively consistent with the prediction of \citet{Becker2022ApJ...939...67B} that increasing luminosity in a radiation-dominated column enhances the radiative deceleration of the plasma flow near the stellar surface, and the associated enhancement of the PdV work done on the radiation. This agreement supports the interpretation that the source has transitioned into a radiation-supported accretion-column regime, allowing the accretion column to develop a more vertically extended structure that may provide the fundamental conditions for triggering internal instabilities.
The CRSF modulation therefore reflects an intrinsic, dynamical reconfiguration of the accretion column geometry on timescales of tens of seconds, linking the emergence of high-energy QPOs to the formation of a radiation-supported column in which rapid geometric restructuring becomes dynamically possible. Given the strong coupling between the QPO and the CRSF component in the 
energy spectrum, we propose a model in which the variability arises from processes intrinsic to the accretion column itself, independent of direct magnetosphere–disk interactions.

\begin{figure*}
    \centering
    \includegraphics[width=0.9\textwidth]{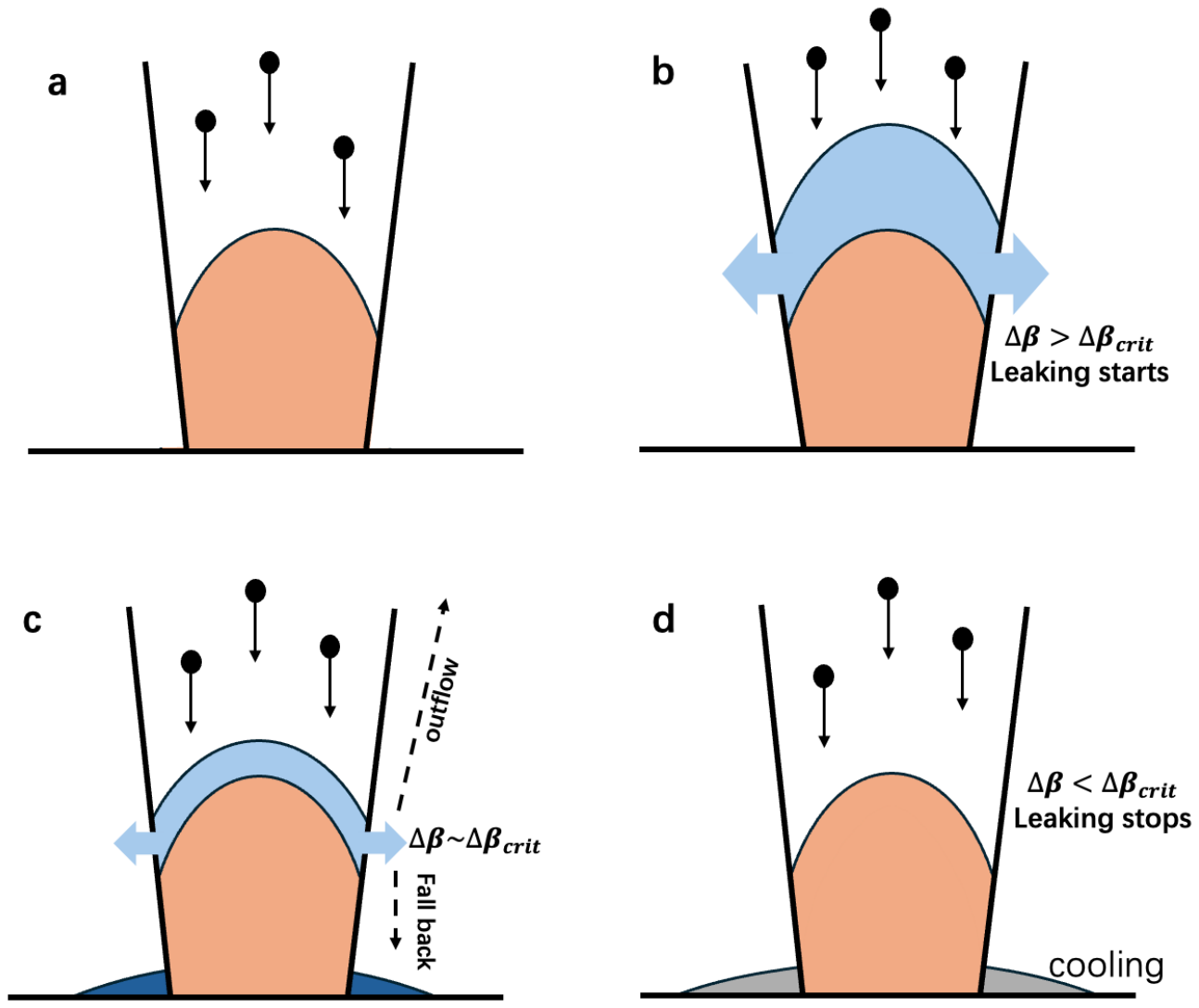}
    \caption{
    Schematic illustration of the accretion column leakage--replenishment cycle. (a) Accreting flow along magnetic field lines accumulates at the magnetic pole. Radiation pressure, thermal pressure, and Coulomb braking decelerate the matter in the post-shock region, gradually building up a vertically extended accretion column. Magnetic pressure supports the column geometry. (b) When the accretion layer $\Delta M$ causes the plasma overpressure at some height to exceed a critical threshold ($\Delta \beta > \Delta \beta_{\text{crit}}$), magnetic confinement is locally disrupted. Part of the plasma leaks sideways into regions of weaker magnetic field near the edge of the polar cap. (c) Sustained leakage reduces the pressure supporting the column, leading to a drop in overpressure. As the leakage continues, depending on the field-line structure (open or closed field lines) and the local radiation field (Eddington limit), the leaked material may either (i) fall back along closed magnetic field lines or (ii) escape along open field lines as a mildly relativistic outflow. (d) As the leakage subsides and the overpressure falls below the critical threshold ($\Delta \beta < \Delta \beta_{\text{crit}}$), accretion begins to replenish the column structure. The cycle then restarts. The fallback material rapidly cools and spreads over the neutron star surface. This feedback process can modulate the height and structure of the accretion column, leading to observable quasi-periodic oscillations (QPOs) in the X-ray light curve and variations in the cyclotron resonance scattering feature (CRSF).
    }
    \label{Schematic}
\end{figure*}

In high accretion rate systems, accretion columns are vertically extended structures supported by thermal and radiation pressure. Numerical simulations \citep{Abolmasov2023MNRAS.524.4148A} show that strong advection and dominant radiation pressure shift the region where thermal pressure exceeds magnetic pressure to a finite height above the neutron star surface. As matter accumulates, the plasma overpressure increases until a critical threshold ($\Delta\beta > \Delta\beta_{\mathrm{crit}}$) is reached. At this stage, interchange (ballooning-type) instabilities \citep{Litwin2001ApJ...553..788L} can be triggered near the column boundary, leading to localized magnetic field deformation and cross-field mass leakage. The leaked material may either accrete laterally along closed field lines and spread over the stellar surface, or escape along open field lines, potentially forming mildly relativistic outflows ($\sim 0.4c$) \citep{Abolmasov2023MNRAS.524.4148A}. The associated mass loss reduces internal pressure (gas and radiation), causing partial column collapse and a decrease in height. Continued accretion then replenishes the column, restarting the cycle. Fig.~\ref{Schematic} illustrates the suggested scenario. This leakage–replenishment process naturally drives oscillations in the column's structure and emission. Variations in column height modulate the local magnetic field strength at the emission site: a taller column emits from regions of weaker magnetic field, producing lower CRSF energies, whereas a collapsed column emits closer to the surface, yielding higher CRSF energies. In this way, the instability acts as a self-regulating mechanism that maintains approximate pressure–field equilibrium while generating quasi-periodic modulation.

In this scenario, the oscillation frequency $\nu_{\rm MDLM}$ can be calculated as follows. The instability condition can be written as
\begin{equation}
\Delta \beta = \frac{\Delta p}{B^2 / 8\pi} > \Delta \beta_{\mathrm{crit}} \sim \frac{8a}{h},
\end{equation}
where $a$ is the lateral radius of the column and 
$h = -\left( \frac{1}{\rho} \frac{d\rho}{dz} \right)^{-1}$
is the density scale height \citep{Litwin2001ApJ...553..788L}. The mass involved in a leakage episode can be estimated by equating the overpressure to the weight of the displaced plasma,
\begin{equation}
\Delta M \approx \frac{S\Delta p}{g}
= \frac{S B^2 \Delta\beta}{8\pi g},
\end{equation}
where $S$ is the cross-sectional area of the column base and $g$ is the surface gravitational acceleration. The corresponding replenishment timescale is
\begin{equation}
t_{\rm r} = \frac{\Delta M}{\dot{M}}
= \frac{S B^2 \Delta \beta}{8\pi g \dot{M}}.
\end{equation} 
For disk accretion with a ring-shaped footprint \citep{Mushtukov2015MNRAS.447.1847M},
\begin{equation}
S \approx (3 \times 10^9\, \mathrm{cm}^2)\,
\Lambda^{-7/8} M_{1.4}^{-13/20}
R_6^{19/10} B_{12}^{-1/2} L_{37}^{2/5}.
\end{equation}
Substituting $S$ yields
\begin{equation}
t_{\rm r} \approx (12\, \mathrm{s}) 
\Delta \beta
\Lambda^{-7/8} 
M_{1.4}^{-13/20} 
R_6^{29/10} 
B_{12}^{3/2} 
L_{37}^{-3/5},
\end{equation}
and therefore
\begin{equation}
\nu_{\rm MDLM} = \frac{1}{t_r}
\propto L_{37}^{3/5},
\end{equation}
in good agreement with the observed relation (Fig.~\ref{QPO_fit}). Adopting representative parameters ($M_{1.4}=1$, $R_6=1$, $B_{12}=5$),
\begin{equation}
\nu_{\rm MDLM}
= 7.45\,\Delta\beta^{-1}
\Lambda^{7/8}
L_{37}^{3/5}
\ \mathrm{(mHz)}.
\end{equation}
Because $\Delta\beta$ and $\Lambda$ are degenerate, taking $\Lambda=0.1$ and $0.5$, the fitting in Fig.~\ref{QPO_fit} yields $\Delta\beta=0.05\pm0.01$ or $0.21\pm0.01$, respectively. These values imply that the vertical extent of the column must significantly exceed its lateral scale. Such conditions are unlikely near the neutron-star crust ($h/a\ll0.13$; \citealt{Litwin2001ApJ...553..788L}), but become plausible in a radiation-dominated column where shocks and radiation-pressure braking elevate the region in which thermal pressure can overcome magnetic confinement \citep{Abolmasov2023MNRAS.524.4148A}. 
Such a scenario is consistent with the threshold luminosity for the emergence of QPOs shown in Fig.~\ref{Long_Short}, where QPOs appear only after the accretion column enters the radiation-supported regime.

Taken together, our results suggest that the appearance of mHz QPOs in 1A~0535+262 marks a dynamically unstable accretion column with a relatively large scale height and a relatively small lateral radius. Within the magnetic field deformation and mass leakage scenario, the observed CRSF modulation arises naturally from a leakage--replenishment cycle that reshapes the column geometry on timescales of tens of seconds. However, for systems with lower accretion rates, those fed by winds, or those with more complex magnetic fields and accretion geometries, this condition would be extremely difficult to achieve. Although this interpretation remains to be verified by future radiation--MHD simulations, it makes a potential observational prediction: for this system, while accreting matter, it also experiences significant mass loss. The intermittent outflow from leakage driven by radiation pressure will introduce a radio emission mechanism to explain the radio emission \citep{Eijnden2022MNRAS.516.4844V}. In addition, plasma that escapes magnetic confinement but remains gravitationally bound may spread over a broader region of the neutron-star surface, effectively enlarging the accretion footprint. This lateral redistribution of matter could introduce asymmetries between the rising and declining phases of an outburst, potentially manifesting as hysteresis or structural differences in the cyclotron-line and continuum evolution \citep{Doroshenko2017MNRAS.466.2143D, Wang2020MNRAS.497.5498W}.

In the supplementary materials, we further evaluated the magnetic disk precession model (MDPM), where QPOs originate from magnetically driven precession of the inner disk. While mHz frequencies can be achieved under certain parameter configurations, the predicted luminosity scaling and the required inner-disk conditions are inconsistent with the observed trend. We also consider the applicability of the model in other sources, such as 1A~1118$-$615, V~0332+53, and 4U~1626$-$67, all of which exhibit persistent QPO signals during their outbursts. The current results suggest that, based on existing observations, 1A~0535+262 is the only known source in which the QPO signal is plausibly driven by intrinsic instabilities within the accretion column itself. However, it is important to emphasize that the origin of QPOs in accreting pulsars is unlikely to be universal. Most sources exhibit markedly different observational properties, and many still lack sufficiently sensitive data or quantitative analyses to draw firm conclusions.

This behaviour emerges in the luminosity regime where radiation pressure becomes dynamically important. 
We therefore expect the proposed column-intrinsic mechanism to operate most readily when several conditions are met: the source is near or above the critical luminosity, a radiation-supported column has formed, magnetic confinement remains strong enough to channel the flow but the column has a large vertical scale height, and the CRSF is measurable with sufficient hard-X-ray signal-to-noise. This interpretation leads to testable predictions. The QPO onset should occur near the transition to radiation-dominated column behaviour; QPO-phase-resolved spectroscopy should reveal coherent CRSF modulation, often anti-correlated with apparent flux; the QPO rms spectrum should peak around the CRSF band; and lower-luminosity systems without strong radiation pressure should be more likely to show disk-boundary-like QPO phenomenology. Although the present result is demonstrated in a single source, the combined use of CRSF measurements and the time–frequency decomposition method (VMD-HHT) establishes a general framework for probing the physical origin of QPOs. Extending this approach to a broader population of accreting pulsars will test whether radiation-induced column instability is common in high-accretion-rate, strongly magnetized systems.

\section{Methods}\label{sec2}

\subsection{Data Selection and Preprocessing}

To probe accretion-column variability on QPO timescales, we require continuous, high signal-to-noise observations with broad energy coverage extending into the cyclotron line regime. The Hard X-ray Modulation Telescope, (Insight-HXMT), was launched on June 15, 2017 \citep{2014SPIE.9144E..21Z,2020SCPMA..63x9502Z}. Its scientific payload consists of three collimated telescopes: the High Energy X-ray Telescope \citep{2020SCPMA..63x9503L} (HE: $5000~\rm cm^2$, 20--250 keV), the Medium Energy X-ray Telescope \citep{2020SCPMA..63x9504C} (ME: $952~\rm cm^2$, 5--30 keV), and the Low Energy X-ray Telescope \citep{2020SCPMA..63x9505C} (LE: $384~\rm cm^2$, 1--10 keV). The corresponding fields of view are $1.6^{\circ}\times6^{\circ}$, $1^{\circ}\times4^{\circ}$, and $1.1^{\circ}\times5.7^{\circ}$ for LE, ME, and HE, respectively. Owing to its large effective area above 30 keV and the absence of pile-up effects even for bright sources, Insight-HXMT is particularly well suited for studying CRSFs and hard X-ray QPOs. 

Insight-HXMT observed 1A~0535+262 from November 6, 2020 (MJD 59159) to December 24, 2020 (MJD 59207), covering the entire giant outburst with a total exposure of $\sim1.91$~Ms. The data were processed using the Insight-HXMT Data Analysis Software (HXMTDAS) v2.05 and calibration model v2.06. 
Standard screening criteria were applied: elevation angle (ELV) $>10^{\circ}$, geomagnetic cutoff rigidity (COR) $>8$~GeV, and pointing offset $<0.04^{\circ}$. 
Data within 300~s of South Atlantic Anomaly passages were excluded. 
When the Crab nebula entered the detector field of view, the affected detector boxes were removed to avoid contamination \citep{Kong2021ApJ...917L..38K}. 

Timing analyses were performed using the \texttt{Stingray} package \citep{Huppenkothen2019ApJ...881...39H} and the \texttt{powspec} task from XRONOS 6.0 \citep{Stella1992ASPC...25..103S} to generate a power density spectrum (PDS) using the background-subtracted light curve. Spectral fitting was carried out with XSPEC v12.14.0h \citep{Arnaud1996ASPC..101...17A}. Background spectra were estimated using the standard tools LEBKGMAP, MEBKGMAP, and HEBKGMAP (v2.0.14), based on the current Insight-HXMT background models \citep{Liao2020a, Guo2020JHEAp, Liao2020b}. To account for calibration uncertainties, systematic errors of 1\% were applied to LE, ME, and HE spectra. Daily exposures were combined using \texttt{addspec} and \texttt{addrmf} to improve statistics. 
Spectra were grouped with \texttt{ftgrouppha} to ensure a minimum of 1000 counts for LE and ME, and 5000 counts for HE. 
Parameter uncertainties were estimated using Markov Chain Monte Carlo (MCMC) simulations with a chain length of $10^{5}$, and are quoted at the 90\% confidence level.

\subsection{Time--Frequency Decomposition Using VMD--HHT}

To isolate variability components on QPO timescales from broadband accretion-rate fluctuations and high-frequency noise, we apply a time--frequency decomposition based on the Hilbert--Huang Transform (HHT) combined with Variational Mode Decomposition (VMD). The HHT is a powerful technique for analyzing non-linear and non-stationary signals \citep{SOUZA2022103292}, and has been widely applied across disciplines including medicine, finance, geophysics, oceanography, and astronomy \citep{Yan4014733, DEZHKAM2023105626, ZHOU201268, DATIG20041783, ORTEGA2009212, Shui2023ApJ...957...84S, Shui2024ApJ...965L...7S, Zhao2024ApJ...961L..42Z}.

Traditional HHT implementations rely on Empirical Mode Decomposition (EMD), which suffers from several limitations, including sensitivity to noise, mode mixing, and a lack of a rigorous mathematical foundation \citep{huang2008review}. 
VMD overcomes these issues by decomposing a signal into a predefined number of intrinsic mode functions (IMFs) with well-defined central frequencies and bandwidths, providing a more stable and physically interpretable decomposition \citep{dragomiretskiy2014variational}. See details in the supplementary materials. Unlike EMD-based methods, the VMD decomposition solves all modes simultaneously through a global variational optimization rather than via sequential sifting. As a result, the extracted QPO component is not biased by mode mixing or by the order of subtraction, enabling a more robust separation between broadband noise and quasi-periodic variability. By comparing the individual IMFs with the power density spectrum (PDS), we find that the characteristic RMS amplitude decreases monotonically from the low-frequency component to the high-frequency components. Accordingly, we adopt a labeling convention in which the IMFs are ordered from low to high characteristic frequency.

Prior to VMD analysis, we subtract the average pulse profile from the light curve to suppress phase-coherent pulsations and focus on aperiodic variability on QPO timescales. The pulse-subtracted HE light curves (25--80~keV) were then decomposed using the \texttt{VmdTransformer} package \citep{dragomiretskiy2014variational, CARVALHO2020102073}.  
Fig.~\ref{PDS_fit} shows that the power density spectrum (PDS) of the pulse-subtracted light curve (black) can be decomposed into five IMFs, which can be classified as the reconstructed contributions from low-frequency noise (blue), the QPO component (red), and high-frequency noise (green), demonstrating that the decomposition accurately reproduces the observed PDS.
In the supplementary material, we present all IMFs obtained from the VMD analysis and discuss their respective physical origins. We have also investigated the CRSF behaviour associated with the other VMD-derived IMFs and find that only the QPO component can exhibit significant, phase-coherent CRSF modulation. Therefore, in this work, we focus exclusively on the QPO component.

\backmatter

\bmhead{Author contributions}

L.K and A.S drafted the main paper with contributions from co-authors. X.D and L.K analysed most of the data from Insight-HXMT and completed data collation, figures and tables. L.J, V.S, A.M, L. D, S.Z, Q.S, SN.Z, H.F, S.S, H.L, P.W and Q.L participated in in-depth model-related discussions. All authors read, commented on and approved the submission of this article.

\bmhead{Competing interests}

The authors declare no competing interests.

\bmhead{supplementary materials}

The Supplementary Figures, Tables and Materials are included after the main manuscript in this compiled PDF.

\bmhead{Acknowledgements}

This work is supported by the National Key R\&D Program of China (2021YFA0718500). Lingda Kong is supported by the Fundamental Research Funds for the Central Universities (010-63263126) and the support provided by the Sino-German (CSC-DAAD) Postdoc Scholarship Program, 2022 (57607866). Long Ji is supported by the National Natural Science Foundation of China under grant No. 12173103. Lorenzo Ducci acknowledges funding from the Deutsche Forschungsgemeinschaft (DFG, German Research Foundation) - Project number 549824807. Sergey S. Tsygankov acknowledges support from the Research Council of Finland Centre of Excellence in Neutron-Star Physics (grant 374064).

\clearpage
\onecolumn
\setcounter{figure}{0}
\setcounter{table}{0}
\renewcommand{\thefigure}{S\arabic{figure}}
\renewcommand{\thetable}{S\arabic{table}}
\renewcommand{\thesection}{S\arabic{section}}
\makeatletter
\renewcommand{\theHfigure}{supp.figure.\arabic{figure}}
\renewcommand{\theHtable}{supp.table.\arabic{table}}
\renewcommand{\theHsection}{supp.section.\arabic{section}}
\renewcommand{\theHsubsection}{supp.subsection.\arabic{section}.\arabic{subsection}}
\makeatother
\setcounter{section}{0}
\begin{center}
{\Large\bfseries Supplementary Materials}\par
\vspace{1em}
{\large Rapid quasi-periodic reconfiguration of the accretion column in pulsar 1A 0535+262}\par
\end{center}
\vspace{1em}

\section{Variational Mode Decomposition and Hilbert Transform Analysis}

The Variational Mode Decomposition (VMD) is a technique used to break down a signal into intrinsic mode functions (IMFs), each representing specific frequency components. This process begins with the construction of an analytic signal for each IMF, expressed as:
\begin{equation}
    s_k(t) = \left[ \delta(t) + \frac{j}{\pi t} \right] * u_k(t),
\end{equation}
where $\delta(t)$ is the Dirac delta function, $u_k(t)$ represents the $k$-th IMF, and $*$ denotes the convolution operator. This step ensures that each IMF retains the necessary information for subsequent analysis.

To isolate distinct frequency bands, each IMF undergoes a frequency-shifting operation using complex modulation:
\begin{equation}
    u_k(t) \cdot e^{-j \omega_k t},
\end{equation}
where $\omega_k$ is the centre frequency of the IMF. The core objective of VMD is to minimize the spectral bandwidth of each IMF, which is achieved by solving the following optimization problem:
\begin{equation}
    \min_{\{u_k\}, \{\omega_k\}} \left\{ \sum_{k=1}^{K} \left\| \partial_t \left[ \left( \delta(t) + \frac{j}{\pi t} \right) * u_k(t) \right] e^{-j \omega_k t} \right\|_2^2 \right\},
\end{equation}
where $\partial_t$ represents the time derivative.

To simplify this constrained optimization problem, a quadratic penalty term and Lagrange multipliers are introduced, transforming it into an unconstrained problem defined by the Lagrangian:


\begin{equation}
\begin{split}
L(u_k, \omega_k, \lambda) ={} & \alpha \sum_{k=1}^{K}
\left\| \partial_t \left[
\left( \delta(t) + \frac{j}{\pi t} \right) * u_k(t)
\right] e^{-j \omega_k t} \right\|_2^2 \\
& + \left\| f(t) - \sum_{k=1}^{K} u_k(t) \right\|_2^2 \\
& + \left\langle \lambda(t),
f(t) - \sum_{k=1}^{K} u_k(t) \right\rangle .
\end{split}
\label{eq:vmd_loss}
\end{equation}aaa

where $f(t)$ is the input signal, $\lambda(t)$ is the Lagrange multiplier, and $\alpha$ is a parameter controlling the bandwidth of the modes.

The Alternating Direction Method of Multipliers (ADMM) is used to iteratively solve for $u_k$ and $\omega_k$. The choice of $\alpha$ and the number of modes $K$ play a significant role in achieving accurate decomposition. In our analysis, $\alpha$ was set between 100 and 500, and $K$ was chosen between 3 and 7 to effectively capture the characteristics of the PDS. For Obs.~1, the best values are $\alpha=100$ and $K=5$. For Obs.~2, the best values are $\alpha=100$ and $K=6$. For Obs.~3, the best values are $\alpha=100$ and $K=7$. Increasing $\alpha$ beyond 500 and $K$ beyond 7 results in a pronounced mismatch between the intrinsic mode functions (IMFs) and the signal's power density spectrum (PDS), indicating a loss of decomposition accuracy.

Once the IMFs are derived, the Hilbert transform is applied to analyze the instantaneous frequency. The Hilbert transform is defined as:
\begin{equation}
    H[IMF_i(t)] = \frac{1}{\pi} P \int \frac{IMF_i(\tau)}{t - \tau} d\tau,
\end{equation}
where $P$ indicates the Cauchy principal value. The analytic signal corresponding to each IMF is constructed as:
\begin{equation}
    z_i(t) = IMF_i(t) + jH[IMF_i(t)] = a_i(t)e^{j\theta_i(t)},
\end{equation}
where $a_i(t)$ represents the instantaneous amplitude, and $\theta_i(t)$ is the instantaneous phase. The instantaneous frequency is then computed as:
\begin{equation}
    \omega_i(t) = \frac{d\theta_i(t)}{dt}.
\end{equation}

\begin{figure*}
    \centering
    \includegraphics[width=0.9\textwidth]{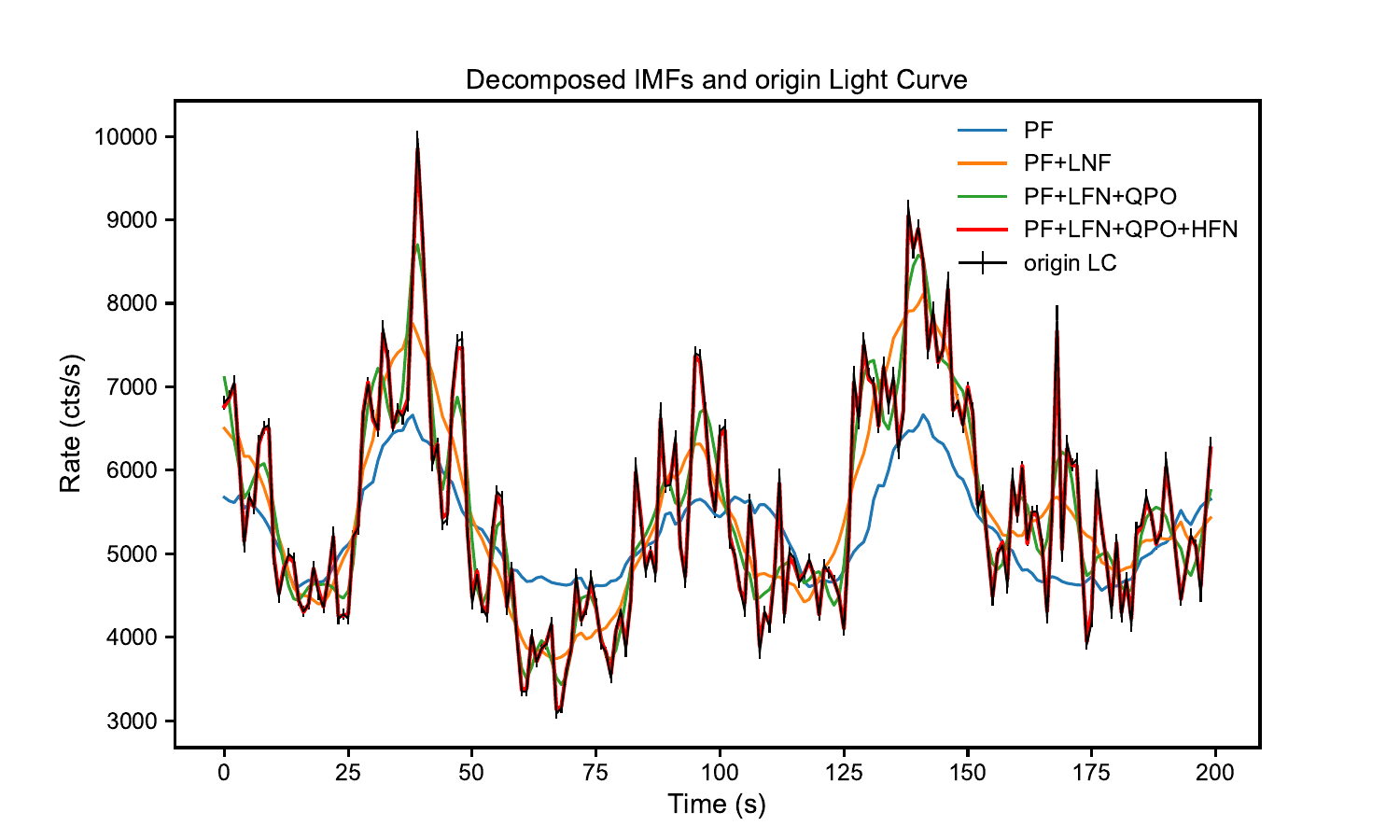}
    \caption{
    The IMFs decomposed from the original light curve using VMD-HHT are shown. The orange line shows that the combination of the average pulse profile and low-frequency noise captures the general outline of the light curve. The green line, which additionally includes the QPO component, fits the short-timescale modulations. Finally, the red line incorporates random narrow peaks, providing a good fit to the full original light curve.
    }
    \label{IMF_fit_LC}
\end{figure*}

This combined VMD-Hilbert transform approach provides a precise method for analyzing non-stationary signals, enabling detailed characterization of dynamic frequency components (see Fig.~\ref{IMF_fit_LC}). The selection of parameters ensures that the decomposition aligns with the signal's physical properties, particularly in applications involving QPO analysis.

\subsection{VMD Decomposition and Physical Interpretation of the Extracted IMFs}

We explicitly present all five intrinsic mode functions (IMFs) obtained from the VMD analysis and discuss their respective physical interpretations.

The light-curve decomposition was performed in two steps. The source exhibits a strong and coherent pulsation with a spin period of approximately 100~s. Because this pulse component has a stable phase and a well-defined profile, it does not require decomposition via VMD. We therefore first folded the original light curve at the spin period to construct the average pulse profile. Since all other variability components—including the QPO—have random phases relative to the spin period, their contributions are averaged out in this folding process. The average pulse profile was then subtracted from the original light curve prior to applying VMD.

After removal of the coherent spin component, the power spectrum of the residual light curve reveals at least three distinct variability components: low-frequency broadband noise, the mHz QPO, and higher-frequency noise. We therefore adopted five IMFs in the VMD decomposition to capture these components in a flexible yet controlled manner.

Among the extracted modes, IMF1 represents the low-frequency broadband noise. Its physical origin is most plausibly associated with stochastic fluctuations arising from disk–magnetosphere interactions, which induce chaotic variations in the instantaneous mass accretion rate. Such fluctuations cause individual pulses to deviate from the average pulse profile and are commonly observed in accreting X-ray pulsars (e.g., \cite{Klochkov2011AA...532A.126K}). Although these variations can modulate the CRSF energy, they follow the established long-term CRSF–luminosity relation and therefore cannot account for the phase-dependent CRSF modulation observed over the QPO cycle.

The higher-frequency modes (IMF3--IMF5) correspond to faster stochastic variability and are interpreted as high-frequency noise components. These may originate from intrinsic processes within the accretion column, such as oscillations of the radiative shock or local thermal–radiative instabilities (e.g., \cite{Zhang2022MNRAS.515.4371Z, Zhang2023MNRAS.520.1421Z}; \cite{Abolmasov2023MNRAS.524.4148A}). These modes do not exhibit coherent phase behaviour and show no systematic correlation with CRSF parameters.

In contrast, the QPO signal is isolated in a distinct IMF that exhibits coherent quasi-periodic behaviour. This component forms the basis of our QPO-phase–resolved spectroscopy and underpins the physical interpretation presented in the main text.

\subsection{CRSF Modulation in Other IMFs}

Here, we further discuss whether the CRSF energy exhibits abnormal modulation associated with IMFs other than the QPO component identified in the VMD decomposition.

The lowest-frequency IMF represents broadband stochastic variability associated with irregular fluctuations in the instantaneous mass accretion rate. Spectrally, this component produces small luminosity variations analogous to pulse-to-pulse accretion-rate changes observed in other accreting X-ray pulsars (e.g., \cite{Klochkov2011AA...532A.126K}). In such cases, the CRSF energy follows the established long-term CRSF--luminosity relation. Consistently, we find that the CRSF variations associated with this IMF are weak and consistent with standard accretion-rate–driven evolution, without exhibiting coherent phase-dependent modulation.

The higher-frequency IMFs (IMF3--IMF5) correspond to rapid stochastic processes with low fractional RMS amplitudes and random phases. Owing to their weak amplitudes and lack of coherent phase structure, these components do not produce systematic spectral changes, and no significant CRSF modulation can be detected when phase-resolved analysis is performed on these modes.

Therefore, among all extracted components, only the QPO IMF exhibits coherent quasi-periodic behaviour accompanied by a significant and systematic modulation of the CRSF energy.

\section{Can the conventional paradigm of long-term CRSF--luminosity evolution be used to explain the coherent modulation of the CRSF during the QPO on short timescales?}

\subsection{Emission Height in the Supercritical Accretion Regime}

\begin{figure*}
    \centering
    \includegraphics[width=0.9\textwidth]
    {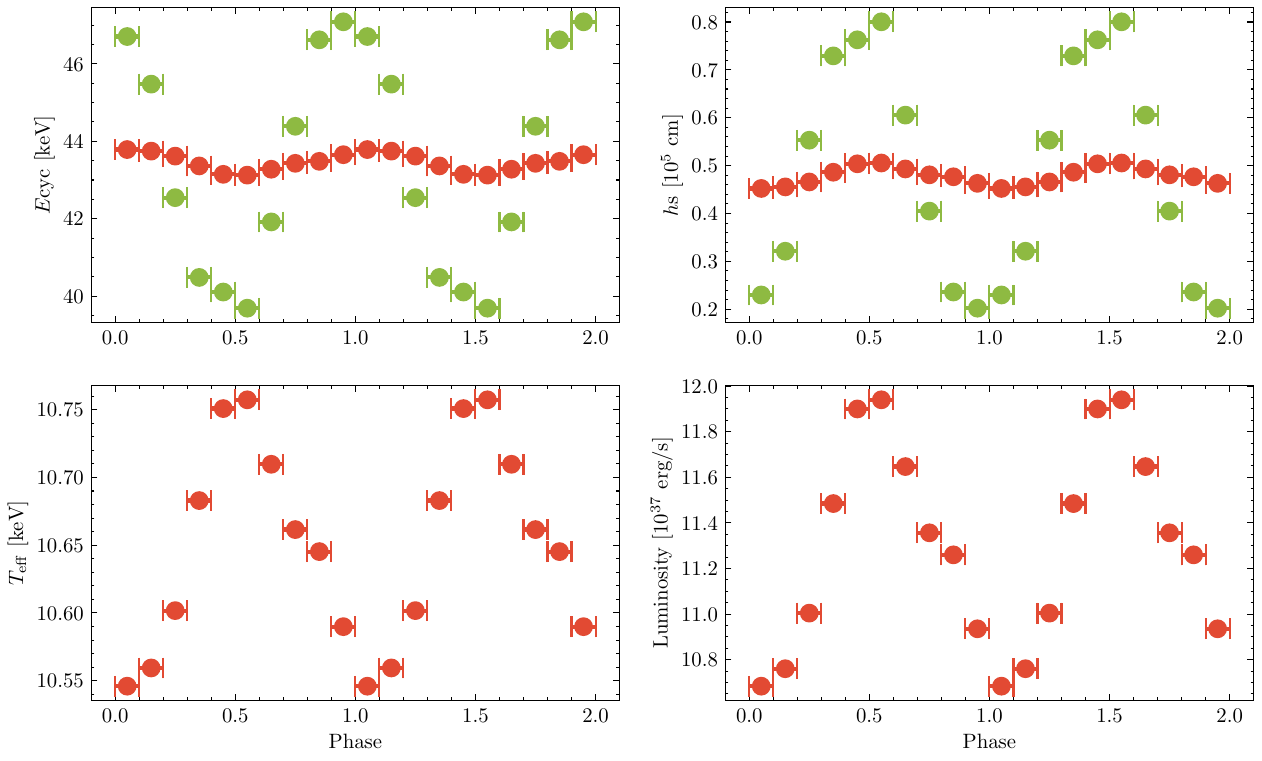}
    \caption{
    The green points of $E_{\rm cyc}$ represent the CRSF energy obtained from spectral fitting on November~20, showing a modulation between 39 and 47~keV. The red points of luminosity and $T_{\rm eff}$ indicate the luminosity from spectral fitting and the effective temperature of the radiation in the post-shock region using Eq.~10. The green points of $h_{s}$ are determined using Eq.~13, representing the observed emission height under the assumption of a CRSF energy of 50~keV at the neutron star's surface. The red points of $h_{s}$ depict the theoretical values calculated from Eq.~14, assuming a supercritical accretion column with typical parameters for accreting pulsars ($\xi=0.018$). Based on this, the theoretical CRSF energy, $E_{\rm cyc}$ (red points), is derived accordingly.
    }
    \label{column_model}
\end{figure*}

Here, we stress our argument that the large energy change produced over approximately 10 seconds during the QPO oscillation cannot be explained by the effect of luminosity changes on the height of the emission site in the accretion column \citep{Becker2012AA...544A.123B}. Applied to the supercritical accretion regime, we calculate the parameter variations implied by our observed CRSF modulation. In Fig.~\ref{column_model}, the red points of $T_{\rm eff}$ show the effective temperature in the post-shock region:
\begin{equation}
    T_{\text{eff}} = 4.35 \times 10^7 \, \text{K}\ \left( \frac{\Lambda}{0.1} \right)^{1/4}\left( \frac{M_{\ast}}{1.4 \, M_\odot} \right)^{-5/56}\left( \frac{R_{\ast}}{10 \, \text{km}} \right)^{-15/56}\left( \frac{B_{\ast}}{10^{12} \, \text{G}} \right)^{1/7}\left( \frac{L_X}{10^{37} \, \text{erg s}^{-1}} \right)^{5/28},
\end{equation}
where
\begin{equation}
    \dot{M} = \frac{L_X \, R_{\ast}}{G \, M_{\ast}},
\end{equation}
\begin{equation}
    B_{\ast} = E_{\rm CRSF}^{\ast} \times 10^{12} / 11.6.
\end{equation}
Here $B_{\ast}$, $M_{\ast}$, and $R_{\ast}$ are the surface magnetic field, mass, and radius of the neutron star, respectively. 
The parameter $\Lambda$, related to the accretion-flow geometry, is set to 0.5. We set the CRSF energy at the surface to $E_{\rm CRSF}^{\ast} = 50$~keV, which is the highest fundamental CRSF energy observed in 1A~0535+262 \citep{Shui2024MNRAS.528.7320S}.
The green points in the top-left panel indicate the fundamental CRSF energy, $E_{\rm cyc}$, derived from the spectral fitting. 
Based on the assumption of a dipole magnetic field, we can calculate the height of the emission site. The red points of $h_s$ are derived from the observed energy centroids:
\begin{equation}
    \frac{E_{\text{cyc}}}{E_{\rm CRSF}^{\ast}} = \left( \frac{R_{\ast} + h_s}{R_{\ast}} \right)^{-3}.
\end{equation}
The height of the emission site in the supercritical regime is:
\begin{equation}
    h_s = 2.28 \times 10^3 \, \text{cm} 
    \left( \frac{\xi}{0.01} \right)
    \left( \frac{M_{\ast}}{1.4 \, M_\odot} \right)^{-1}
    \left( \frac{R_{\ast}}{10 \, \text{km}} \right)
    \left( \frac{L_X}{10^{37} \, \text{erg s}^{-1}} \right),
\end{equation}
where
\begin{equation}
    \xi \equiv \frac{v_{\text{eff}}}{v} \sim \frac{1}{\mathcal{M}_\infty^2} \ \text{(hydrostatic approximation)}.
\end{equation}
The parameter $\xi$ is defined as the ratio of the effective velocity for photon transport, $v_{\rm eff}$, to the flow velocity, $v$. The Mach number $\mathcal{M}_{\infty}$ represents the incident (upstream) Mach number of the flow with respect to the radiation sound speed \citep{Becker1998ApJ...498..790B}.
However, the precise value of $\xi$ is difficult to estimate, as it depends on the complexity of radiation hydrodynamics, including factors such as the radiation sound speed, photon advection and diffusion, the velocity profile in the accretion column, and other related processes \citep{Basko1976MNRAS.175..395B, Becker1998ApJ...498..790B}.

In our calculations, we assume that the cyclotron line energy values are modulated around their mean value, as is the emission height. Based on this assumption, we find that $\xi$ needs to vary around 0.018.
This value is consistent with the reasonable range $10^{-3}$--$10^{-2}$, as a low effective velocity tends to result in a compact emission region in supercritical sources \citep{Becker1998ApJ...498..790B}.
By substituting a constant $\xi = 0.018$, the red points show the calculated variations in the emission height and the cyclotron line energy according to changes in the accretion rate.

Therefore, it is evident that an accretion column under the hydrostatic assumption cannot rely solely on variations in the accretion rate to explain the observed changes in the CRSF energy; the variations caused by accretion-rate changes are significantly smaller than the amplitude of the observed modulation.

\subsection{Relativistic Motion of the Infalling Plasma}

Alternative models, for instance, relativistic motion of accretion plasma toward the neutron star surface, could redshift the CRSF via the Doppler effect. Higher radiation pressure would decelerate the infalling plasma near the surface, leading to a higher observed CRSF energy \citep{Mushtukov2015MNRAS.454.2714M}. Yet, this mechanism predicts a positive correlation between CRSF energy and flux, which is inconsistent with our observations.

\subsection{Reflection on the Neutron Star Surface}

The reflection model posits that cyclotron resonance scattering features (CRSFs) in X-ray pulsars originate not within the accretion column, but from the neutron star surface where column radiation is reflected \citep{Poutanen2013ApJ...777..115P}. At high luminosities, a taller column and emission site illuminate a broader surface region with a lower average magnetic field strength than at the poles. This naturally produces the observed negative correlation between CRSF centroid energy and luminosity, successfully explaining the behaviour in V~0332+53. However, Monte Carlo simulations reveal that incorporating photons reflected from all illuminated latitudes significantly smears the CRSF feature, effectively erasing the predicted anti-correlation \citep{Kylafis2021A&A...655A..39K}. Furthermore, in their simulations, the model predicts a convex CRSF energy--luminosity dependence, inconsistent with the distinctly concave evolution observed in 1A~0535+262, indicating that the reflection model alone is insufficient \citep{2024A&A...689A..75L}.

Even so, we still explored whether reflection could explain CRSF energy oscillations correlated with the QPO phase, distinct from the long-term trend. Since most of the emission originates from a region of characteristic scale $\Delta h$ situated above the neutron star surface at height $h$, making the average magnetic field strength more sensitive to the irradiation area requires $\Delta h/h$ to be close to 0, as shown in Table~1 of \cite{Poutanen2013ApJ...777..115P}. In this scenario, we assume that the height of the emission site, i.e., the height of the radiation-dominated shock, can somehow be modulated on the QPO timescale. This change in height alters the illuminated fraction of the neutron star surface, thereby determining the average sampled magnetic field (see Fig.~1 in \cite{Poutanen2013ApJ...777..115P}). As a result, this scenario also predicts a negative relation between the CRSF energy and the QPO flux. However, even under such assumptions, a cyclotron-line energy variation of about 10\% ($\sim4$~keV) would still require a substantial change in the emission height. This again cannot be accounted for by accretion-rate variations alone and therefore calls for an additional mechanism to explain such modulation.

\section{Dependence on the Adopted CRSF Line Model}

In the main text, the CRSF is modeled using a Gaussian absorption profile (\texttt{gabs}), which provides a convenient phenomenological description of the line feature. To assess whether the choice of line profile affects our results, we performed an alternative set of fits using the physically motivated \texttt{cyclabs} model, while keeping the underlying continuum model unchanged.

\begin{figure*}
    \centering
    \includegraphics[width=0.9\textwidth]{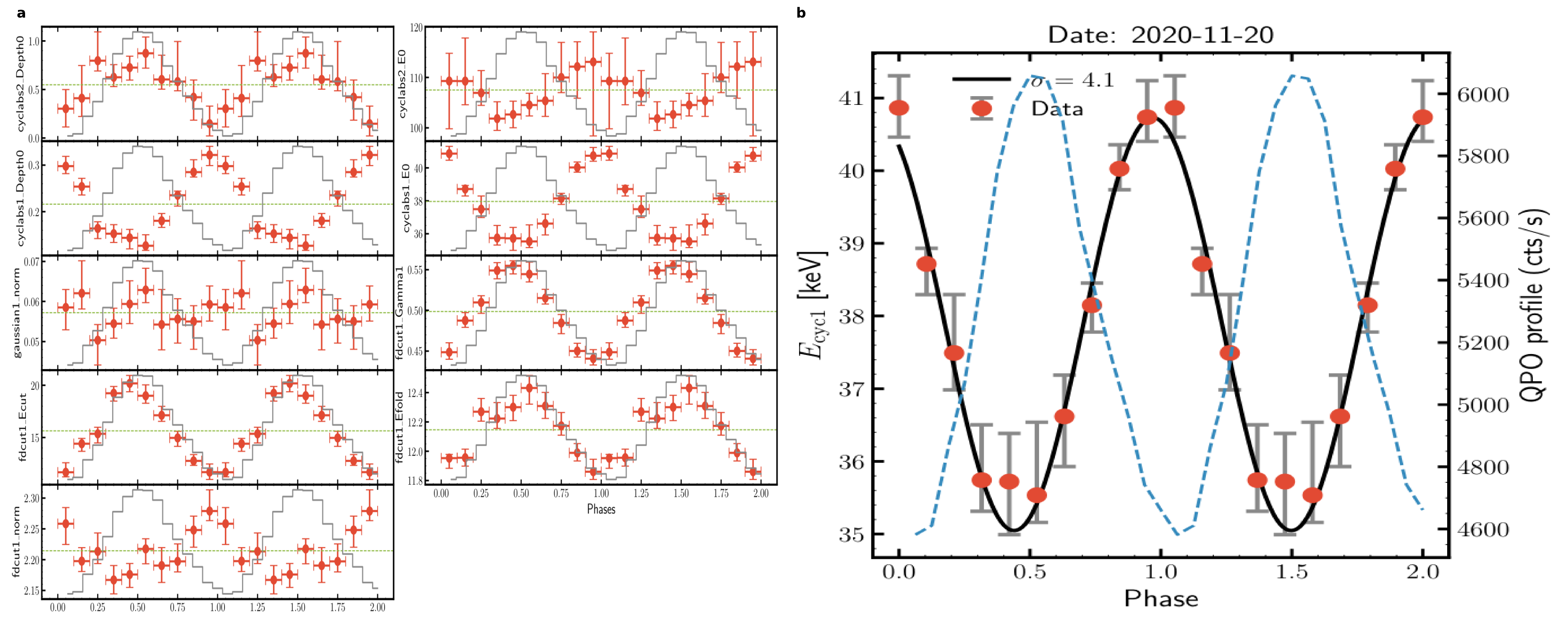}
    \caption{The parameters of fit with \texttt{TBabs}$\times$\texttt{cyclabs}$\times$\texttt{cyclabs}$\times$(\texttt{gaussian}+\texttt{fdcut}) model. And the modulation of CRSF energy.}
    \label{cyclabs_fit}
\end{figure*}

In Fig.~\ref{cyclabs_fit}, we find that the overall fit quality remains comparable between the two line prescriptions. Most importantly, the phase-dependent modulation of the CRSF centroid energy over the QPO cycle persists when using \texttt{cyclabs}. The amplitude and phase behaviour of the CRSF modulation are consistent within statistical uncertainties between the two models.

While \texttt{cyclabs} can account for asymmetric line shapes expected from resonant scattering in a magnetized plasma, the inferred CRSF energy evolution and its anti-correlation with QPO flux are not sensitive to the adopted line profile. We therefore retain the \texttt{gabs} model in the main text for clarity and consistency with previous works, and conclude that our primary results are robust against reasonable choices of CRSF line modeling.

\section{Physical Diversity of Low-Frequency QPOs in Accreting Pulsars}

Low-frequency QPOs (LFQPOs) represent a more ubiquitous phenomenon, manifesting as millihertz broadband peaks in power density spectra. Their characteristics exhibit marked diversity: transient LFQPOs emerge during specific accretion states in sources like RX~J0440.9+4431 \citep{LiRXJ0440_2024}, IGR~J19294+1816 \citep{Raman2021AstroSatDO}, V0332+53 \citep{Qu2005DiscoveryON, CaballeroGarcia2015ActivityFT, Manikantan2024EnergyDO}, GX~301-2 \citep{DevasiaGX301m2_2011}, and Cen~X-3 \citep{Raichur2008ApJ...685.1109R, Aftab2019XRayRT, Liu2022MNRAS.516.5579L}, whereas persistent signals typify 1A~0535+262 \citep{Finger1996ApJ...459..288F, Ma2022MNRAS.517.1988M}, 1A~1118-615 \citep{Nespoli2011A&A...526A...7N}, and 4U~1626-67 \citep{Jain2009SpectralAT, Sharma20234U1R}. Intriguingly, LFQPO amplitudes display heterogeneous energy dependencies—positive, negative, or uncorrelated with photon energy \citep{Manikantan2024EnergyDO}. 1A~0535+262 exemplifies this complexity, exhibiting long-lived LFQPOs exclusively above 25~keV, with amplitude peaking near 60~keV \citep{Ma2022MNRAS.517.1988M}. This variability in duration and spectral behaviour strongly suggests multiple physical origins for LFQPOs, a premise demanding systematic investigation. Except for 1A~0535+262, which is discussed in the main text, we further reconsider the origins of QPOs in 1A~1118-615, V~0332+53, and 4U~1626-67, all of which exhibit persistent QPO signals during their outbursts. 

Among them, 1A~1118-61 exhibits a CRSF at $\sim55$~keV, and its QPO frequency evolution with luminosity is consistent with the Keplerian scaling $\nu_{\rm QPO} \propto L_{\rm X}^{3/7}$. The QPO rms remains at $\sim5\%$ and shows no clear correlation with flux, while it decreases with increasing energy above 4~keV and disappears above 15~keV \citep{Nespoli2011A&A...526A...7N}. These properties are consistent with a QPO origin associated with obscuration by inhomogeneities at the inner disk–magnetosphere boundary. Considering that the historical peak flux of this source is relatively low and that the CRSF energy is independent of luminosity \citep{Maitra2012MNRAS.420.2307M}, the radiation pressure in this source is weak, and the accretion column may not be fully developed. Therefore, it does not fall within the scope of the MDLM model.

For V~0332+53, although its CRSF behaviour indicates that the accretion column has been clearly established \citep{Doroshenko2017MNRAS.466.2143D}, $\nu_{\rm QPO}$ remains constant even when the flux changes by a factor of 4.5. The QPO rms remains constant as a function of photon energy up to 10~keV and drops beyond 10~keV \citep{Qu2005DiscoveryON}. These properties cannot be explained by luminosity-related models such as KFM/BFM or MDLM. Furthermore, this source also exhibits various other types of QPOs, which may indicate multiple physical origins of QPOs in this system.

For 4U~1626-67, whose QPOs appear only during several torque-reversal stages, QPO phase-resolved spectral analysis using the HHT method reveals that the QPO modulation is caused by accretion-rate variability \citep{Zhou2025ApJ...994...46Z}.

\section{Canonical and generalized KFM/BFM scalings}
\label{sec:generalized_kfm_bfm}

The mHz QPOs are often interpreted in terms of the Keplerian-frequency model \citep{Klis1997ASSL..218..121V} (KFM) or the beat-frequency model \citep{Alpar1985Natur.316..239A} (BFM). In the KFM, the QPOs arise from the periodic obscuration of X-rays by inhomogeneities in the inner disk at the Keplerian frequency [i.e., $\nu_{\rm QPO}=\nu_{\rm K}(r_{\rm in})$]. In the BFM, blobs of matter at the inhomogeneous inner boundary of the accretion disk, orbiting with the Keplerian frequency, are accreted onto the neutron star at a rate modulated by the rotating magnetic field. As a result, the observed QPO frequency represents the beat between the Keplerian frequency $\nu_{\rm K}$ at the inner disk radius $r_{\rm in}$ and the neutron-star spin frequency $\nu_{\rm s}$ [i.e., $\nu_{\rm QPO}=\nu_{\rm K}(r_{\rm in})-\nu_{\rm s}$].

For the canonical dipolar magnetospheric scaling, the inner disk radius is written as
\begin{equation}
    R_{\rm m}=\Lambda R_{\rm A}, \qquad
    R_{\rm A}=\left(\frac{\mu^4}{2GM\dot M^2}\right)^{1/7},
\end{equation}
where $\mu=BR^3$ and $\Lambda$ is assumed to be independent of luminosity. If $L_X\propto\dot M$, the Keplerian frequency at the inner disk radius is
\begin{equation}
    \nu_{\rm K}(R_{\rm m}) = \frac{1}{2\pi}\left(\frac{GM}{R_{\rm m}^3}\right)^{1/2}=(530\ \rm mHz)\ \Lambda^{-3/2} B_{12}^{-6/7} M_{1.4}^{2/7} R_{6}^{-15/7} L_{37}^{3/7}
\end{equation}
, where $\Lambda$ is a constant that depends on the accretion-flow geometry, $B_{12}$ is the magnetic field in units of $10^{12}$~G, $R_{6}$ is the neutron-star radius in units of $10^{6}$~cm, $M_{1.4}$ is the neutron-star mass in units of 1.4 times the solar mass.
With typical parameters of 1A~0535+262, $\Lambda=0.5$, $B_{12}=4.7$, $M_{1.4}=1$, $L_{37}\equiv\alpha\beta \frac{GM\dot{M}}{R}/10^{37}=11$, $\alpha=1$, and $\beta=1$, we estimate the Keplerian frequency $\nu_{\rm K}=538$~mHz or the beat frequency $\sim520$~mHz, both of which are higher than the observed QPO frequency of $\sim94$~mHz. Matching the observed normalization would require $\Lambda\sim2.4$ for KFM-like parameters, which is larger than the usual disk-accretion expectation of $\Lambda\lesssim1$ \citep{Finger1996ApJ...459..288F}. This would imply that the QPO might form at a radius much further out rather than in the inner region of the accretion disk, or that the flow in the inner region of the disk is sub-Keplerian \citep{Camero-Arranz2012ApJ...754...20C}.

\begin{figure*}
    \centering
    \includegraphics[width=0.95\textwidth]{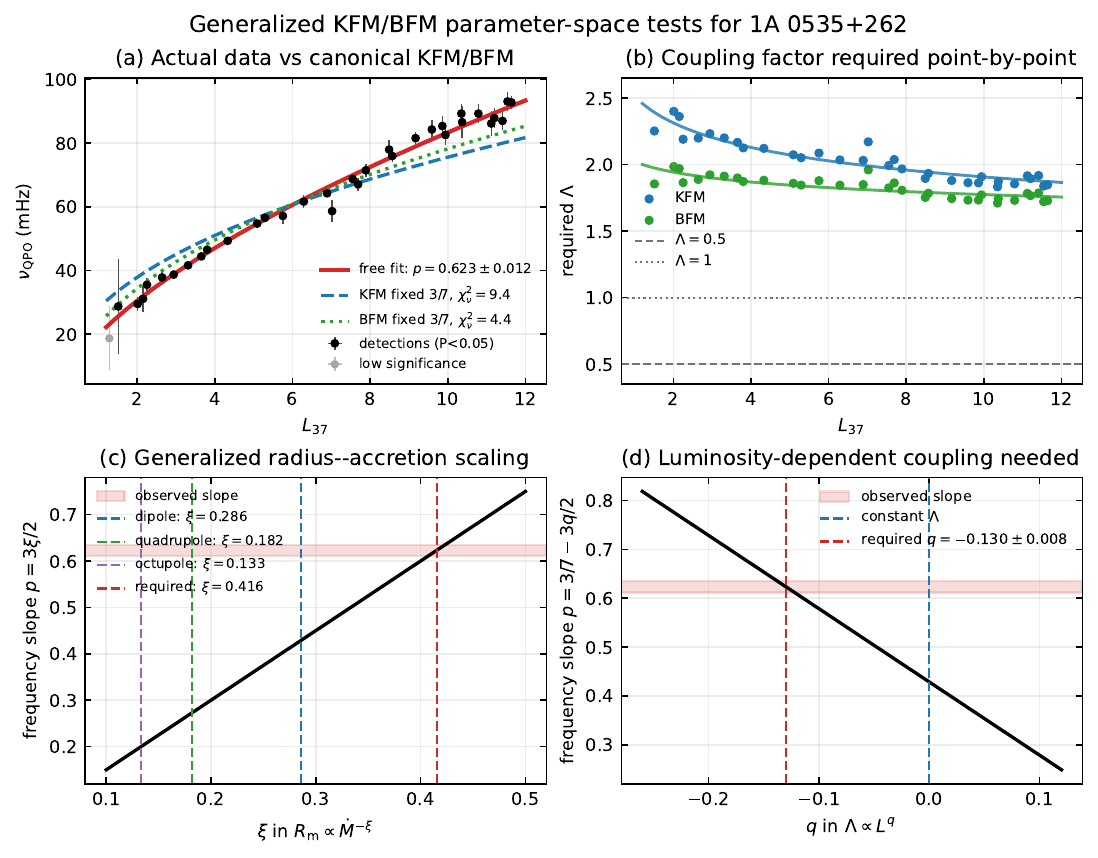}
    \caption{Generalized KFM/BFM parameter-space tests for 1A~0535+262. (a) The measured $\nu_{\rm QPO}$--$L_X$ relation is well described by a free power law with $p=0.623\pm0.012$, whereas the canonical fixed-coupling KFM and BFM prescriptions based on the $3/7$ magnetospheric scaling give substantially poorer fits. (b) Inverting the KFM/BFM relations point-by-point gives the effective coupling factor $\Lambda$ required to match the data for the fiducial source parameters. The required values are systematically above the usual disk-accretion expectation of $\Lambda\lesssim1$. (c) If $R_{\rm m}\propto\dot M^{-\xi}$, the observed slope requires $\xi\simeq0.416$, steeper than the canonical dipole value $2/7$; higher multipoles move in the opposite direction by producing shallower scalings. (d) If the canonical dipolar radius exponent is retained but the coupling factor varies as $\Lambda\propto L^q$, matching the observed slope requires $q=-0.130\pm0.008$, i.e. a systematic decrease of the effective coupling factor with luminosity.}
    \label{fig:generalized_kfm_bfm}
\end{figure*}

For 1A~0535+262, the observed relation, $\nu_{\rm QPO}\propto L_X^p$ with $p=0.623\pm0.012$, is significantly steeper than this canonical $3/7$ dependence. However, the $3/7$ exponent is not a unique prediction of all disk--magnetosphere frequency models. It depends on the assumptions of a dipolar field, standard pressure balance, a thin disk, and a luminosity-independent coupling factor. Magnetic threading, non-stationary disk--magnetosphere coupling, instabilities, and radiation-pressure effects can in principle alter the relation between $R_{\rm m}$ and $\dot M$ \citep{Wang1995ApJ...449L.153W,Romanova2002ApJ...578..420R,Romanova2003ApJ...595.1009R,DAngelo2010MNRAS.406.1208D}. Figure~\ref{fig:generalized_kfm_bfm} summarizes the corresponding parameter-space tests using the measured QPO--luminosity data.

To quantify the required departure from the canonical case, we write a generalized inner-radius scaling as
\begin{equation}
    R_{\rm m}\propto \dot M^{-\xi}.
\end{equation}
Then
\begin{equation}
    \nu_{\rm K}\propto R_{\rm m}^{-3/2}\propto \dot M^{3\xi/2}.
\end{equation}
If the observation-averaged luminosity traces the secular mass-accretion rate, $L_X\propto\dot M$, the observed slope gives
\begin{equation}
    \xi=\frac{2p}{3}=0.416\pm0.008.
\end{equation}
This is larger than the canonical dipole value $\xi=2/7=0.286$. Higher multipoles give even shallower scalings; for example, a quadrupolar field gives $\xi=2/11=0.182$, and an octupolar field gives $\xi=2/15=0.133$. Therefore, multipolar structure alone does not naturally explain the observed steeper slope if the QPO is identified with a Keplerian frequency.

Another possibility is to retain the dipolar Alfv\'en-radius dependence but allow the coupling factor to vary with luminosity, $\Lambda\propto L_X^q$. In this case
\begin{equation}
    \nu_{\rm K}\propto \Lambda^{-3/2}L_X^{3/7}\propto L_X^{3/7-3q/2},
\end{equation}
so that
\begin{equation}
    q=\frac{3/7-p}{3/2}=-0.130\pm0.008.
\end{equation}
Thus the observed slope can be reproduced mathematically if the effective disk--magnetosphere coupling factor decreases systematically as luminosity increases. Such a prescription is possible but introduces additional physics that is not independently constrained by the present data. It also remains highly degenerate with beaming and bolometric-correction effects.

We therefore conclude that the observations disfavor the canonical KFM/BFM implementations with standard dipolar magnetospheric scaling and a luminosity-independent coupling factor. More generalized disk--magnetosphere frequency models cannot be excluded by the $\nu_{\rm QPO}$--$L_X$ slope alone. Nevertheless, such models must also explain the CRSF-centered hard-X-ray rms peak and the coherent QPO-phase anti-correlation between apparent flux and CRSF centroid energy, which directly implicate the cyclotron-line-forming region.

\section{Magnetic Disk Precession Model (MDPM)}

The magnetic disk precession model (MDPM) attributes QPOs to magnetically driven warping or precession of the inner accretion disk, induced by torques between the neutron star’s magnetic field and surface currents in a misaligned disk \citep{Lai1999ApJ...524.1030L, Shirakawa2002ApJ...565.1134S}. In this framework, the inner disk becomes tilted and precesses around the stellar spin axis, leading to quasi-periodic modulation whose frequency is expected to correlate positively with the mass accretion rate.

While periodic occultation of the neutron star by the precessing disk could modulate X-ray flux, this predicts a broadband signal inconsistent with the observed high-energy exclusivity of QPOs. Instead, we propose the possibility that disk precession periodically alters the structure at the disk–magnetosphere boundary, thereby modulating the geometry of the accretion channel feeding the column. Crucially, the penetration depth of the disk into the magnetosphere ($\delta\simeq 2H$, where $H$ is the disk scale height \citep{Lai2014EPJWC..6401001L}) governs the accretion channel's cross-section. Precession-induced variations in this depth cause periodic changes in the channel's width and, consequently, the accretion column's height and lateral extent. A wider channel produces a shorter, broader column with a stronger effective global magnetic field, while a narrower channel yields a taller, thinner column with a weaker field. This geometric modulation could drive the observed periodic shifts in cyclotron line energy.

From the MDPM model, the QPO frequency is given by \cite{Shirakawa2002ApJ...565.1134S}: 
$\nu_{\mathrm{QPO}}=(0.83\ \mathrm{mHz})\mathscr{K}\dot{M}_{17}^{0.71}\mathscr{J}_{\mathrm{in}}^{-0.7}$, 
where $\mathscr{K}=A(\frac{\Lambda}{0.5})^{-4.93}\sin^2\theta\mu_{30}^{-0.81}\alpha_{-1}^{0.85}M_{1.4}^{0.18}$, 
$A=0.3$--$0.85$ (depending on details of the disk structure), $\theta$ is the angle between the magnetic dipole axis and the spin axis, and $\alpha_{-1}$ is the dimensionless viscosity parameter in units of 0.1. The value of $\mathscr{J}_{\mathrm{in}}$ depends on the physical conditions at the inner edge of the disk. 

Among these sources, 1A~1118-61 exhibits a CRSF at $\sim55$~keV, and its QPO frequency evolution with luminosity is consistent with the Keplerian scaling $\nu_{\rm QPO}\propto L_{\rm X}^{3/7}$. The QPO rms remains at $\sim5\%$ and shows no clear correlation with flux, while it decreases with increasing energy above 4~keV and disappears above 15~keV \citep{Nespoli2011A&A...526A...7N}. These properties are consistent with a QPO origin associated with inhomogeneities at the inner disk–magnetosphere boundary. Considering that the historical peak flux of this source is relatively low and that the CRSF energy is independent of luminosity \citep{Maitra2012MNRAS.420.2307M}, the radiation pressure in this source is weak, and the accretion column may not be fully developed. Therefore, it does not fall within the scope of the MDLM model.

For V~0332+53, although its CRSF behaviour indicates that the accretion column has been clearly established \citep{Doroshenko2017MNRAS.466.2143D}, $\nu_{\rm QPO}$ remains constant even when the flux changes by a factor of 4.5. The QPO rms remains constant as a function of photon energy up to 10~keV and drops beyond 10~keV \citep{Qu2005DiscoveryON}. These properties cannot be explained by luminosity-related models such as KFM/BFM or MDLM. Furthermore, this source also exhibits various other types of QPOs, which may indicate multiple physical origins of QPOs in this system.

For 4U~1626-67, whose QPOs appear only during several torque-reversal stages, QPO phase-resolved spectral analysis using the HHT method reveals that the QPO modulation is caused by accretion-rate variability.

The minimum value of $\mathscr{J}_{\mathrm{in}}$ is given by
\[
\mathscr{J}_{\mathrm{min}} \simeq 3.1 \times 10^{-4} \alpha_{-1}^{9/8} M_{1.4}^{-7/16} \dot{M}_{17}^{1/4} \left( \frac{r_{\mathrm{in}}}{10^8\, \mathrm{cm}} \right)^{1/16},
\]
where the inner radius $r_{\mathrm{in}}$ of the disk is given by the magnetospheric radius
\[
R_{\rm m}=\Lambda \left( \frac{\mu^4}{G M \dot{M}^2} \right)^{1/7}
= (3.4 \times 10^8\, \mathrm{cm})\, \Lambda\, \mu_{30}^{4/7} M_{1.4}^{-1/7} \dot{M}_{17}^{-2/7}.
\]

The MDPM can produce QPOs with frequencies of order $1\,\mathrm{mHz}$; larger values of $\nu_{\mathrm{QPO}}$ would require $\mathscr{J}_{\mathrm{in}} \ll 1$ (corresponding to low surface density). With $\Lambda=0.5$, $\alpha_{-1}=1$, $\mu_{30}=5$, and $\dot{M}_{17}=6.1$, we estimate 
$\nu_{\rm QPO}=(0.81\ \mathrm{mHz})A\sin^2 \theta \mathscr{J}_{\mathrm{in}}^{-0.7}$. 
For a 90~mHz QPO, if $A=0.85$, $\mathscr{J}_{\mathrm{in}}$ should be $\sim9.5\times10^{-4}$, which satisfies the condition $\mathscr{J}_{\mathrm{min}}\simeq 5\times10^{-4}$. 

Under the assumption that $\mathscr{J}_{\mathrm{in}}$ and $\mathscr{K}$ remain unchanged during the outburst, $\nu_{\rm QPO}\propto L^{0.71}$, which increases faster than the observed relation ($\nu_{\rm QPO}\propto L^{0.62}$). Even when allowing for an accretion-rate dependence in $\mathscr{J}_{\mathrm{in}}$, the large number of poorly constrained parameters entering the model introduces substantial degeneracy, rendering a quantitative assessment highly uncertain. A comprehensive exploration of this parameter space is therefore beyond the scope of the present work.

\section{Pulse Profiles in Different QPO Phases}

\begin{figure*}
    \centering
    \includegraphics[width=0.9\textwidth]{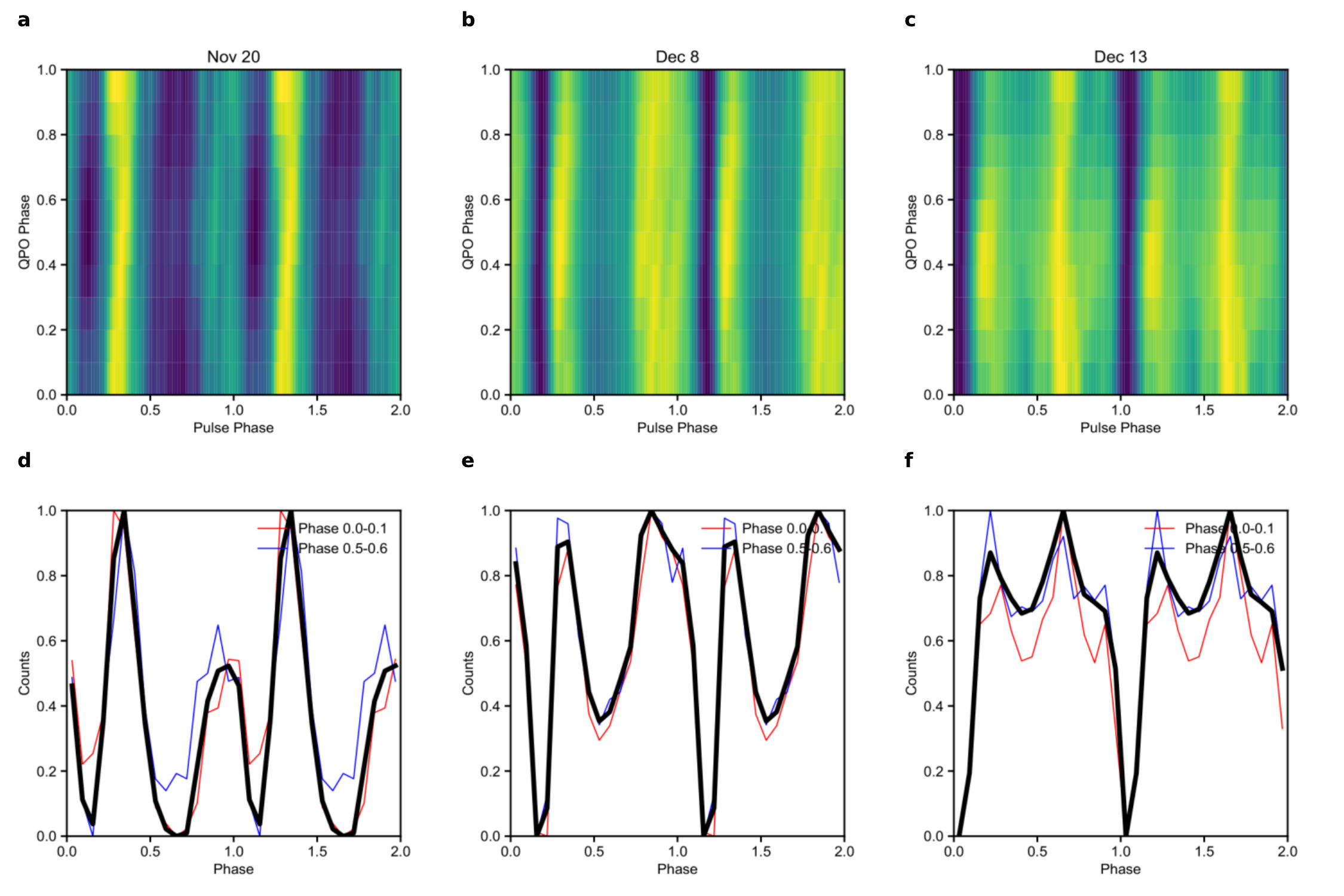}
    \caption{
    Panels (a--c) present the two-dimensional QPO-resolved pulse profiles in 25--80~keV for the three observations. The colors show the amplitudes of the pulse profiles, normalized to the [0,1] range. Panels (d--f) show the pulse profiles corresponding to the QPO phase average (black line), phases 0.0--0.1 (red line), and 0.5--0.6 (blue line). The shapes of the pulse profiles exhibit slight variations.
    }
    \label{sub_pulse}
\end{figure*}

To probe potential geometric reorganization of the accretion column, we compared pulse profiles across quasi-periodic oscillation (QPO) phases. Fig.~\ref{sub_pulse}a--c displays the evolution of the 25--80~keV pulse morphology with QPO phase for three observations, normalized to the [0,1] range to isolate intrinsic shape variations. The overall profile structure remains statistically invariant across phases. Further quantification is presented in Fig.~\ref{sub_pulse}d--f, contrasting phase-averaged profiles with those from phases 0.0--0.1 and 0.5--0.6. The peak amplitude ratios exhibit minor phase-dependent modulation, and the pulse morphology shows slight structural reorganization. This indicates that the column geometry changes during QPO cycles.

\section{Modulation of the Harmonic CRSF Centroid Energy}

\begin{figure*}
    \centering
    \includegraphics[width=0.9\textwidth]{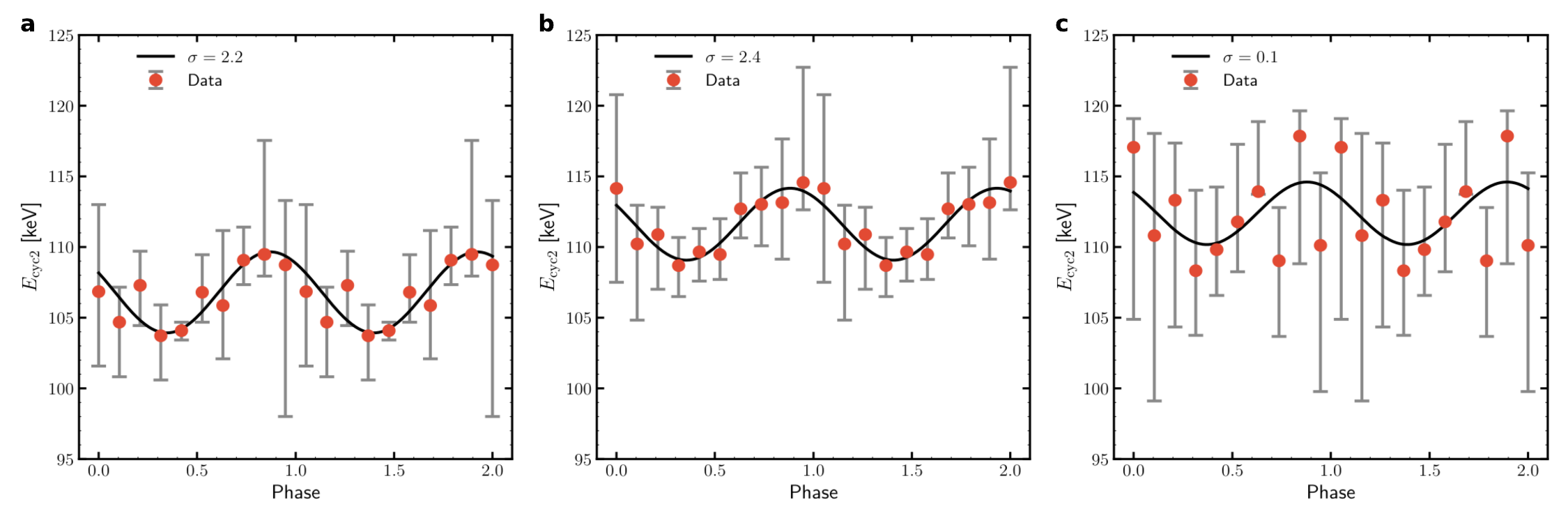}
    \caption{
    QPO Phase Modulation of the Harmonic CRSF Centroid Energy. Panels (a), (b), and (c) represent the results from observations on November~20, December~8, and December~13, 2020, respectively. The red points show the harmonic CRSF centroid energy obtained from spectral fitting at different QPO phases, while the black curves represent the best-fitting sinusoidal modulation. The significance of each modulation is indicated in the figure. This analysis suggests an anti-correlation between the CRSF energy and QPO phase.
    }
    \label{CRSF_harmonic_modulation}
\end{figure*}

In addition to the modulation of the fundamental CRSF energy, the harmonic CRSF energy also shows modulation with QPO phase. In Fig.~\ref{CRSF_harmonic_modulation}, we present the $E_{\rm cyc2}$ values obtained from spectral fitting across different QPO phases. We note that the significance level of this modulation is $\leq 2\sigma$, making it statistically insignificant and preventing us from ruling out the possibility that it arises from purely statistical fluctuations.

\section{Supplementary Data}\label{secED}


\begin{table*}
\centering
\caption{List of Insight--HXMT observations for QPO searching and spectral analysis}
\label{tab:qpo_obs}
\begin{tabular}{cccccc}
\hline
Observation Date & Observation Time & $L_{\mathrm{2-150}}$ & $\nu_{\rm QPO}$ & F-test & HE Exp.\\
(YYYYMMDD) & (UTC) & ($10^{37}\,\mathrm{erg\,s^{-1}}$) & (mHz) & (P-value) & (ks)\\
\hline
20201110 & 00:26:36--07:03:21 & $2.15\pm0.02$ & $31.1 \pm 4.0$ & 1.478e-2 (2.20$\sigma$) & 5.72\\
20201114 & 07:49:43--21:26:33 & $7.02\pm0.01$ & $58.6 \pm 3.5$ & 8.791e-8 (5.33$\sigma$) & 11.35\\
20201115 & 00:37:20--22:52:47 & $8.49\pm0.01$ & $77.9 \pm 3.0$ & 7.369e-10 (6.06$\sigma$) & 15.56\\
20201116 & 05:56:48--22:43:39 & $9.86\pm0.01$ & $85.3 \pm 3.0$ & 1.280e-10 (6.33$\sigma$) & 15.80\\
20201117 & 01:54:26--20:59:06 & $10.79\pm0.01$ & $89.2 \pm 3.0$ & 1.212e-12 (7.03$\sigma$) & 13.10 \\
20201118 & 00:09:52--20:49:57 & $11.54\pm0.01$ & $93.0 \pm 3.0$ & 5.551e-9 (5.75$\sigma$) & 14.38 \\
20201119 & 00:00:44--22:16:23 & $11.41\pm0.01$ & $86.9 \pm 3.0$ & 3.635e-9 (5.83$\sigma$) & 12.76\\
20201120 & 01:27:03--20:31:47 & $11.64\pm0.01$ & $92.7 \pm 2.0$ & 2.590e-11 (6.61$\sigma$) & 12.35\\
20201121 & 05:12:10--21:58:10 & $11.20\pm0.01$ & $87.8 \pm 3.0$ & 2.292e-10 (6.27$\sigma$) & 15.40\\
20201122 & 01:08:58--23:24:45 & $11.12\pm0.01$ & $86.1 \pm 4.0$ & 3.124e-10 6.24$\sigma$ & 14.76\\
20201124 & 16:45:29--23:07:10 & $10.37\pm0.02$ & $86.5 \pm 5.0$ & 3.338e-6 (4.53$\sigma$) & 7.78\\
20201125 & 02:17:54--21:23:09 & $10.35\pm0.01$ & $89.2 \pm 3.0$ & 7.385e-9 (5.67$\sigma$) & 13.24\\
20201126 & 00:33:53--22:50:11 & $9.94\pm0.01$ & $82.5 \pm 3.0$ & 6.333e-10 (6.10$\sigma$) & 13.70\\
20201127 & 05:54:16--22:42:03 & $9.59\pm0.01$ & $84.2 \pm 3.0$ & 2.750e-6 (4.56$\sigma$) & 11.47\\
20201128 & 01:52:52--20:58:45 & $9.17\pm0.01$ & $81.5 \pm 2.0$ & 7.111e-10 (6.07$\sigma$) & 13.90\\
20201129 & 00:09:36--22:26:52 & $8.57\pm0.01$ & $75.9 \pm 2.0$ & 2.106e-7 (5.06$\sigma$) & 14.20\\
20201130 & 07:04:36--23:55:39 & $7.89\pm0.01$ & $71.4 \pm 2.0$ & 1.606e-11 (6.65$\sigma$) & 14.87\\
20201201 & 03:06:40--22:14:06 & $7.69\pm0.01$ & $67.0 \pm 2.0$ & 7.711e-10 (6.05$\sigma$) & 14.80\\
20201202 & 02:11:32--20:33:38 & $7.55\pm0.01$ & $68.7 \pm 2.0$ & 4.160e-8 (5.36$\sigma$) & 17.20\\
20201203 & 05:06:55--21:09:49 & $6.89\pm0.01$ & $64.2 \pm 1.0$ & 3.132e-10 (6.24$\sigma$) & 17.77\\
20201204 & 00:14:19--20:59:03 & $6.29\pm0.01$ & $61.6 \pm 2.0$ & 6.364e-12 (6.77$\sigma$) & 15.02\\
20201205 & 00:04:24--19:18:23 & $5.75\pm0.01$ & $57.1 \pm 2.5$ & 1.141e-7 (5.2$\sigma$) & 15.89\\
20201206 & 06:25:59--22:07:25 & $5.29\pm0.01$ & $56.5 \pm 1.0$ & 2.203e-13 (7.23$\sigma$) & 21.06\\
20201207 & 01:22:44--21:57:28 & $5.09\pm0.01$ & $54.7 \pm 1.0$ & 1.145e-13 (7.34$\sigma$) & 20.22\\
20201208 & 01:14:04--21:28:15 & $4.33\pm0.01$ & $49.3 \pm 1.0$ & 2.102e-13 (7.25$\sigma$) & 21.36\\
20201209 & 04:31:43--18:06:55 & $3.80\pm0.01$ & $46.5 \pm 1.0$ & 7.679e-8 (5.34$\sigma$) & 12.55\\
20201210 & 00:28:16--22:42:41 & $3.65\pm0.01$ & $44.4 \pm 1.0$ & 2.286e-8 (5.49$\sigma$) & 15.05\\
20201211 & 01:53:30--17:46:44 & $3.31\pm0.01$ & $41.6 \pm 1.0$ & 2.318e-7 (5.05$\sigma$) & 11.96\\
20201212 & 00:08:14--23:58:17 & $2.94\pm0.01$ & $38.7 \pm 1.5$ & 9.637e-10 (6.00$\sigma$) & 15.99\\
20201213 & 03:09:07--22:13:19 & $2.64\pm0.01$ & $37.8 \pm 2.0$ & 7.854e-5 (3.80$\sigma$) & 11.68\\
20201214 & 01:24:01--20:28:15 & $2.25\pm0.01$ & $35.5 \pm 2.0$ & 1.326e-4 (3.64$\sigma$) & 9.23\\
20201215 & 05:11:30--21:53:56 & $2.01\pm0.01$ & $29.5 \pm 2.1$ & 1.958e-3 (2.88$\sigma$) & 11.83\\
20201216 & 01:04:39--23:19:32 & $1.51\pm0.01$ & $28.7 \pm 15$ & 7.064e-3 (2.45$\sigma$) & 12.82\\
20201217 & 05:41:06--21:34:42 & $1.28\pm0.01$ & $18.7\pm 10$ & 0.266 (0.60$\sigma$) & 11.98 \\
20201218 & 06:18:40--23:00:22 & $1.30\pm0.01$ & ... & ... & 9.53 \\
20201219 & 02:11:13--21:15:35 & $1.06\pm0.01$ & ... & ... & 9.80 \\
20201221 & 05:50:46--22:31:50 & $0.53\pm0.01$ & ... & ... & 8.60 \\
20201222 & 01:42:42--20:47:09 & $0.41\pm0.01$ & ... & ... & 7.89 \\
20201223 & 03:08:39--22:13:00 & $0.212\pm0.004$ & ... & ... & 5.79 \\
20201224 & 01:23:53--20:28:23 & $0.145\pm0.004$ & ... & ... & 4.45\\
\hline
\end{tabular}
\end{table*}

\begin{table*}[ptbptbptb]
    \begin{center}
\caption{Parameters of the spectral fitting on November 20 by \texttt{fdcut} model}
    \begin{tabular}{cccccccccccc}
\hline
\hline
& Parameters & Average & Phase 0.0-0.1 & Phase 0.5-0.6 
\\
\hline
tbabs & $n_{\rm H}\ (10^{22}\ \rm cm^{-2})$ & $0.44$ (fixed) & $0.44$ (fixed) & $0.44$ (fixed)
\\
gabs1 & $E_{\rm cyc1}$ (keV) & $43.5_{-0.4}^{+0.3}$ & $46.7_{-0.4}^{+0.4}$ & $39.7_{-0.5}^{+0.6}$
\\
& $\sigma_{\rm cyc1}$ (keV) (line width) & $10.0_{-0.5}^{+1.1}$ & $12.5_{-3.0}^{+3.4}$ & $10.2_{-0.4}^{+1.8}$
\\
& $d_{\rm cyc1}$ (keV) (line deep) & $5.0_{-1.0}^{+1.4}$ & $9.5_{-0.5}^{+1.9}$ & $3.2_{-0.7}^{+0.9}$
\\
gabs2 & $E_{\rm cyc2}$ (keV) & $111_{-1}^{+6}$ & $107_{-5}^{+6}$ & $107_{-2}^{+3}$
\\
& $\sigma_{\rm cyc2}$ (keV) (line width) & $11_{-1}^{+4}$ & $10$ (fixed) & $10$ (fixed)
\\
& $d_{\rm cyc2}$ (keV) (line deep) & $17_{-3}^{+6}$ & $7_{-3}^{+3}$ & $16_{-2}^{+2}$
\\
gaussian & $E_{\rm Fe}$ (keV) & $6.57_{-0.01}^{+0.02}$ & $6.6$ (fixed) & $6.6$ (fixed)
\\
& $\sigma_{\rm Fe}$ (keV) & $0.37_{-0.01}^{+0.07}$ & $0.3$ (fixed) & $0.3$ (fixed)
\\
& norm ($10^{-2}$)& $6.6_{-0.2}^{+0.8}$ & $6.0_{-0.4}^{+0.5}$ & $6.2_{-0.8}^{+0.3}$
\\
fdcut & $\Gamma$ & $0.50_{-0.01}^{+0.01}$ & $0.47_{-0.01}^{+0.01}$ & $0.55_{-0.01}^{+0.01}$
\\
& $E_{\rm cut}$ (keV) & $16.5_{-0.1}^{+1.5}$ & $14.7_{-0.6}^{+1.3}$ & $20.1_{-1.1}^{+1.3}$
\\
& $E_{\rm fold}$ (keV) & $11.9_{-0.1}^{+0.1}$ & $11.8_{-0.1}^{+0.1}$ & $12.1_{-0.2}^{+0.1}$
\\
& norm & $2.1_{-0.1}^{+0.1}$ & $2.1_{-0.1}^{+0.1}$ & $2.2_{-0.1}^{+0.1}$
\\
\hline
constant & ME & $1.09_{-0.01}^{+0.01}$ & $1.06_{-0.01}^{+0.01}$ & $1.07_{-0.01}^{+0.02}$
\\
constant & HE & $1.09_{-0.01}^{+0.01}$ & $1.01_{-0.01}^{+0.01}$ & $1.06_{-0.02}^{+0.01}$
\\
unabsorbed Flux & $F_{2-150}$ & $2.29_{-0.01}^{+0.01}$ & $2.15_{-0.01}^{+0.01}$ & $2.49_{-0.01}^{+0.01}$
\\
($10^{-7}$ erg cm$^{-2}$ s$^{-1}$) & & &
\\
unabsorbed Luminosity & $L_{2-150}$ & $11.0_{-0.1}^{+0.1}$ & $10.3_{-0.1}^{+0.1}$ & $11.9_{-0.1}^{+0.1}$
\\
($10^{37}$ erg s$^{-1}$)
\\
Fitting & $\chi^{2}$/d.o.f & 1452/1471 & 724/757 & 781/768
\\
\hline
\hline
    \end{tabular}
    \label{spectral_paras_20201120_fdcut}
\begin{list}{}{}
    \item[Note:]{
    Uncertainties are quoted at the 90\% confidence level, derived from a Markov chain Monte Carlo (MCMC) analysis with a chain length of 10,000. During spectral fitting, systematic errors of 1\% were included for the LE (2--10~keV), ME (10--30~keV), and HE (30--150~keV) bands, respectively. The unabsorbed flux in the 2--150~keV band was computed using the convolution model \texttt{cflux}, considering only the emission from the source: \texttt{TBabs$\times$gabs$\times$gabs$\times$cflux$\otimes$(gaussian+fdcut)}. The corresponding unabsorbed luminosity $L_{2--150}$ was calculated as $4\pi D^2 \times F_{2--150}$, where the source distance $D$ is 2~kpc. The width of the harmonic CRSF $\sigma_{\rm cyc2}$ is fixed at 10 because it could not be well constrained. The $n_{\rm H}$ of the ISM absorption, the line energy $E_{\rm Fe}$, the width $\sigma_{\rm Fe}$, and the normalization of the iron emission line are also fixed.
    }
\end{list}
    \end{center}
\end{table*}

\begin{table*}[ptbptbptb]
    \begin{center}
\caption{Parameters of the spectral fitting on December 8 by \texttt{fdcut} model}
    \begin{tabular}{cccccccccccc}
\hline
\hline
& Parameters & Average & Phase 0.0-0.1 & Phase 0.5-0.6 
\\
\hline
tbabs & $n_{\rm H}\ (10^{22}\ \rm cm^{-2})$ & $0.44$ (fixed) & $0.44$ (fixed) & $0.44$ (fixed)
\\
gabs1 & $E_{\rm cyc1}$ (keV) & $46.5_{-0.3}^{+0.1}$ & $49.2_{-0.3}^{+0.8}$ & $45.5_{-0.8}^{+0.9}$
\\
& $\sigma_{\rm cyc1}$ (keV) (line width) & $11_{-1}^{+2}$ & $11_{-1}^{+2}$ & $8_{-1}^{+2}$
\\
& $d_{\rm cyc1}$ (keV) (line deep) & $8_{-1}^{+1}$ & $12_{-1}^{+1}$ & $2_{-1}^{+1}$
\\
gabs2 & $E_{\rm cyc2}$ (keV) & $112_{-2}^{+4}$ & $114_{-7}^{+7}$ & $109_{-2}^{+3}$
\\
& $\sigma_{\rm cyc2}$ (keV) (line width) & $9_{-1}^{+4}$ & $10$ (fixed) & $10$ (fixed)
\\
& $d_{\rm cyc2}$ (keV) (line deep) & $26_{-5}^{+10}$ & $13_{-6}^{+9}$ & $24_{-4}^{+8}$
\\
gaussian & $E_{\rm Fe}$ (keV) & $6.54_{-0.01}^{+0.04}$ & $6.6$ (fixed) & $6.6$ (fixed)
\\
& $\sigma_{\rm Fe}$ (keV) & $0.19_{-0.01}^{+0.07}$ & $0.3$ (fixed) & $0.3$ (fixed)
\\
& norm ($10^{-2}$)& $1.4_{-0.1}^{+0.1}$ & $1.5_{-0.1}^{+0.1}$ & $1.5_{-0.4}^{+0.4}$
\\
fdcut & $\Gamma$ & $0.54_{-0.01}^{+0.01}$ & $0.64_{-0.01}^{+0.01}$ & $0.63_{-0.02}^{+0.01}$
\\
& $E_{\rm cut}$ (keV) & $7.0_{-2.5}^{+1.1}$ & $9.1_{-1.5}^{+1.6}$ & $7.3_{-1.4}^{+1.3}$
\\
& $E_{\rm fold}$ (keV) & $14.1_{-0.1}^{+0.2}$ & $14.0_{-0.1}^{+0.2}$ & $15.3_{-0.2}^{+0.1}$
\\
& norm & $1.3_{-0.1}^{+0.1}$ & $1.5_{-0.1}^{+0.1}$ & $1.6_{-0.1}^{+0.1}$
\\
\hline
constant & ME & $1.00_{-0.01}^{+0.01}$ & $0.97_{-0.01}^{+0.01}$ & $1.01_{-0.01}^{+0.01}$
\\
constant & HE & $1.01_{-0.01}^{+0.01}$ & $0.95_{-0.01}^{+0.01}$ & $1.01_{-0.01}^{+0.01}$
\\
unabsorbed Flux & $F_{2-150}$ & $0.90_{-0.01}^{+0.01}$ & $0.82_{-0.01}^{+0.01}$ & $0.98_{-0.01}^{+0.01}$ 
\\
($10^{-7}$ erg cm$^{-2}$ s$^{-1}$)
\\
unabsorbed Luminosity & $L_{2-150}$ & $4.3_{-0.1}^{+0.1}$ & $3.9_{-0.1}^{+0.1}$ & $4.7_{-0.1}^{+0.1}$
\\
($10^{37}$ erg s$^{-1}$)
\\
Fitting & $\chi^{2}$/d.o.f & 1454/1406 & 724/757 & 781/768
\\
\hline
\hline
    \end{tabular}
    \label{spectral_paras_20201208_fdcut}
\begin{list}{}{}
    \item[Note]{: Following the notes in Table~\ref{spectral_paras_20201120_fdcut}.}
\end{list}
    \end{center}
\end{table*}

\begin{table*}[ptbptbptb]
    \begin{center}
\caption{Parameters of the spectral fitting on December 13 by \texttt{fdcut} model}
    \begin{tabular}{cccccccccccc}
\hline
\hline
& Parameters & Average & Phase 0.0-0.1 & Phase 0.5-0.6 
\\
\hline
tbabs & $n_{\rm H}\ (10^{22}\ \rm cm^{-2})$ & $0.44$ (fixed) & $0.44$ (fixed) & $0.44$ (fixed)
\\
gabs1 & $E_{\rm cyc1}$ (keV) & $47_{-1}^{+1}$ & $49_{-1}^{+1}$ & $45_{-1}^{+1}$
\\
& $\sigma_{\rm cyc1}$ (keV) (line width) & $12_{-1}^{+2}$ & $12_{-1}^{+2}$ & $13_{-1}^{+4}$
\\
& $d_{\rm cyc1}$ (keV) (line deep) & $11_{-1}^{+1}$ & $16_{-1}^{+1}$ & $8_{-1}^{+2}$
\\
gabs2 & $E_{\rm cyc2}$ (keV) & $108_{-1}^{+2}$ & $117_{-12}^{+2}$ & $112_{-4}^{+5}$
\\
& $\sigma_{\rm cyc2}$ (keV) (line width) & $4_{-1}^{+4}$ & $10$ (fixed) & $10$ (fixed)
\\
& $d_{\rm cyc2}$ (keV) (line deep) & $23_{-6}^{+8}$ & $7_{-4}^{+4}$ & $21_{-5}^{+13}$
\\
gaussian & $E_{\rm Fe}$ (keV) & $6.51_{-0.05}^{+0.02}$ & $6.6$ (fixed) & $6.6$ (fixed)
\\
& $\sigma_{\rm Fe}$ (keV) & $0.12_{-0.07}^{+0.03}$ & $0.3$ (fixed) & $0.3$ (fixed)
\\
& norm ($10^{-2}$)& $0.49_{-0.07}^{+0.05}$ & $0.18_{-0.06}^{+0.40}$ & $0.55_{-0.28}^{+0.32}$
\\
fdcut & $\Gamma$ & $0.61_{-0.01}^{+0.01}$ & $0.60_{-0.01}^{+0.03}$ & $0.59_{-0.02}^{+0.01}$
\\
& $E_{\rm cut}$ (keV) & $\leq$ 2 & 0 (fixed) & 0 (fixed)
\\
& $E_{\rm fold}$ (keV) & $15.3_{-0.1}^{+0.1}$ & $14.8_{-0.1}^{+0.4}$ & $16.4_{-0.17}^{+0.11}$
\\
& norm & $1.11_{-0.01}^{+0.07}$ & $1.19_{-0.01}^{+0.05}$ & $1.16_{-0.02}^{+0.02}$
\\
\hline
constant & ME & $0.94_{-0.01}^{+0.01}$ & $0.94_{-0.01}^{+0.01}$ & $0.95_{-0.01}^{+0.01}$
\\
constant & HE & $0.97_{-0.01}^{+0.01}$ & $0.96_{-0.01}^{+0.01}$ & $0.99_{-0.01}^{+0.01}$
\\
unabsorbed Flux & $F_{2-150}$ & $0.54_{-0.01}^{+0.01}$ & $0.48_{-0.01}^{+0.01}$ & $0.59_{-0.01}^{+0.01}$
\\
($10^{-7}$ erg cm$^{-2}$ s$^{-1}$)
\\
unabsorbed Luminosity & $L_{2-150}$ & $2.6_{-0.1}^{+0.1}$ & $2.3_{-0.1}^{+0.1}$ & $2.8_{-0.1}^{+0.1}$
\\
($10^{37}$ erg s$^{-1}$)
\\
Fitting & $\chi^{2}$/d.o.f & 1331/1400 & 398/435 & 781/768
\\
\hline
\hline
    \end{tabular}
    \label{spectral_paras_20201213_fdcut}
\begin{list}{}{}
    \item[Note]{: Following the notes in Table~\ref{spectral_paras_20201120_fdcut}.}
\end{list}
    \end{center}
\end{table*}

\begin{table*}[ptbptbptb]
    \begin{center}
\caption{Parameters of the spectral fitting on November 20 by \texttt{compmag} model}
    \begin{tabular}{cccccccccccc}
\hline
\hline
& Parameters & Average & Phase 0.0-0.1 & Phase 0.5-0.6 
\\
\hline
tbabs & $n_{\rm H}\ (10^{22}\ \rm cm^{-2})$ & $0.44$ (fixed) & $0.44$ (fixed) & $0.44$ (fixed)
\\
gabs1 & $E_{\rm cyc1}$ (keV) & $42.3_{-0.1}^{+0.2}$ & $45.8_{-0.1}^{+0.1}$ & $37.2_{-0.2}^{+0.3}$
\\
& $\sigma_{\rm cyc1}$ (keV) (line width) & $8_{-1}^{+1}$ & $10_{-1}^{+2}$ & $13_{-2}^{+6}$
\\
& $d_{\rm cyc1}$ (keV) (line deep) & $3_{-1}^{+1}$ &  $6_{-1}^{+1}$ & $2_{-1}^{+1}$
\\
gabs2 & $E_{\rm cyc2}$ (keV) & $111_{-1}^{+1} $ & $106_{-4}^{+8}$ & $115_{-7}^{+2}$
\\
& $\sigma_{\rm cyc2}$ (keV) (line width) & $12_{-1}^{+1}$ & $8_{-2}^{+5}$ & $18_{-1}^{+3}$
\\
& $d_{\rm cyc2}$ (keV) (line deep) & $26_{-3}^{+4}$ & $10_{-3}^{+5}$ & $46_{-17}^{+9}$
\\
bbodyrad & kT (keV) & $0.49_{-0.02}^{+0.01}$ & $0.49$ (fixed) & $0.49$ (fixed)
\\
& norm & $5281_{-226}^{+881}$ & $5281_{-486}^{+539}$ & $6645_{-121}^{+44}$
\\
gaussian & $E_{\rm Fe}$ (keV) & $6.58_{-0.01}^{+0.01}$ & $6.6$ (fixed) & $6.6$ (fixed)
\\
& $\sigma_{\rm Fe}$ (keV) & $0.31_{-0.01}^{+0.07}$ & $0.31$ (fixed) & $0.31$ (fixed)
\\
& norm ($10^{-2}$)& $5.6_{-0.1}^{+0.1}$ & $5.6$ (fixed) & $5.6$ (fixed)
\\
compmag & $kT_{\rm bb}$ (keV) & $1.40_{-0.02}^{+0.01}$ & $1.40$ (fixed) & $1.40$ (fixed)
\\
(A=1, betaflag=1) & $kT_{\rm e}$ (keV) & $4.9_{-0.2}^{+0.1}$ & $4.7_{-0.1}^{+0.1}$ & $5.2_{-0.1}^{+0.1}$
\\
& $\tau$ & $0.9_{-0.1}^{+0.1}$ & $0.9$ (fixed) & $0.9$ (fixed)
\\
& $\eta$ & $0.87_{-0.02}^{+0.02}$ & $0.87$ (fixed) & $0.87$ (fixed) 
\\
& $\beta_0$ & $0.19_{-0.01}^{+0.01}$ & $0.19$ (fixed) & $0.19$ (fixed)
\\
& $r_0$ & $0.31_{-0.01}^{+0.01}$ & $0.31$ (fixed) & $0.31$ (fixed)
\\
& norm & $3162_{-43}^{+111}$ & $3160_{-77}^{+58}$ & $3387_{-28}^{+26}$
\\
\hline
constant & ME & $1.08_{-0.01}^{+0.02}$ & $1.06_{-0.01}^{+0.01}$ & $1.07_{-0.01}^{+0.01}$
\\
constant & HE & $1.08_{-0.02}^{+0.02}$ & $1.05_{-0.01}^{+0.01}$ & $1.04_{-0.01}^{+0.02}$
\\
Fitting & $\chi^{2}$/d.o.f & 1433/1471 & 799/808 & 753/821
\\
\hline
\hline
    \end{tabular}
    \label{spectral_paras_20201120_compmag}
\begin{list}{}{}
    \item[Note:]{
    Following the notes in Table~\ref{spectral_paras_20201120_fdcut}, we note that for the \texttt{compmag} model, the number of free parameters exceeds that of the observational constraints. We therefore fixed the seed photon blackbody temperature $kT_{\rm bb}$, the optical depth $\tau$, the velocity profile index $\eta$, the terminal velocity at the neutron star surface $\beta_0$, and the radius of the accretion column $r_0$ to the values obtained from the average spectrum during the QPO phase-resolved analysis. The blackbody temperature $kT$ is also fixed, as it is independent of the QPO phases.
    }
\end{list}
    \end{center}
\end{table*}

\begin{table*}[ptbptbptb]
    \begin{center}
\caption{Parameters of the spectral fitting on December 8 by \texttt{compmag} model}
    \begin{tabular}{cccccccccccc}
\hline
\hline
& Parameters & Average & Phase 0.0-0.1 & Phase 0.5-0.6 
\\
\hline
tbabs & $n_{\rm H}\ (10^{22}\ \rm cm^{-2})$ & $0.44$ (fixed) & $0.44$ (fixed) & $0.44$ (fixed)
\\
gabs1 & $E_{\rm cyc1}$ (keV) & $46.5_{-0.3}^{+0.1}$ & $48.5_{-0.5}^{+0.7}$ & $42.3_{-3.1}^{+1.0}$
\\
& $\sigma_{\rm cyc1}$ (keV) (line width) & $12_{-1}^{+2}$ & $12_{-1}^{+2}$ & $14_{-3}^{+1}$
\\
& $d_{\rm cyc1}$ (keV) (line deep) & $7.6_{-0.5}^{+0.3}$ & $12_{-1}^{+2}$ & $5_{-1}^{+1}$
\\
gabs2 & $E_{\rm cyc2}$ (keV) & $113_{-2}^{+1}$ & $109_{-1}^{+11}$ & $110_{-7}^{+3}$
\\
& $\sigma_{\rm cyc2}$ (keV) (line width) & $10_{-1}^{+1}$ & $6_{-1}^{+8}$ & $10_{-3}^{+3}$
\\
& $d_{\rm cyc2}$ (keV) (line deep) & $31_{-4}^{+2}$ & $12_{-1}^{+16}$ & $25_{-11}^{+5}$
\\
bbodyrad & kT (keV) & $0.37_{-0.01}^{+0.01}$ & $0.37$ (fixed) & $0.37$ (fixed)
\\
& norm & $6519_{-159}^{+72}$ & $12631_{-977}^{+148}$ & $12587_{-378}^{+414}$
\\
gaussian & $E_{\rm Fe}$ (keV) & $6.57_{-0.01}^{+0.01}$ & $6.6$ (fixed) & $6.6$ (fixed)
\\
& $\sigma_{\rm Fe}$ (keV) & $0.18_{-0.02}^{+0.01}$ & $0.2$ (fixed) & $0.2$ (fixed)
\\
& norm ($10^{-2}$)& $0.2_{-0.1}^{+0.1}$ & $0.2$ (fixed) & $0.2$ (fixed)
\\
compmag & $kT_{\rm bb}$ (keV) & $1.35_{-0.01}^{+0.03}$ & $1.35$ (fixed) & $1.35$ (fixed)
\\
(A=1, betaflag=1) & $kT_{\rm e}$ (keV) & $3.64_{-0.03}^{+0.13}$ & $3.36_{-0.03}^{+0.04}$ & $3.93_{-0.03}^{+0.01}$
\\
& $\tau$ & $0.78_{-0.02}^{+0.02}$ & $0.78$ (fixed) & $0.78$ (fixed)
\\
& $\eta$ & $0.62_{-0.01}^{+0.01}$ & $0.62$ (fixed) & $0.62$ (fixed) 
\\
& $\beta_0$ & $0.24_{-0.01}^{+0.01}$ & $0.24$ (fixed) & $0.24$ (fixed)
\\
& $r_0$ & $0.21_{-0.02}^{+0.02}$ & $0.21$ (fixed) & $0.21$ (fixed)
\\
& norm & $1501_{-74}^{+40}$ & $1578_{-8}^{+11}$ & $1689_{-14}^{+3}$
\\
\hline
constant & ME & $1.05_{-0.01}^{+0.01}$ & $0.99_{-0.02}^{+0.01}$ & $1.01_{-0.01}^{+0.01}$
\\
constant & HE & $1.07_{-0.01}^{+0.01}$ & $1.00_{-0.02}^{+0.02}$ & $1.03_{-0.01}^{+0.01}$
\\
Fitting & $\chi^{2}$/d.o.f & 1387/1401 & 689/682 & 683/652
\\
\hline
\hline
    \end{tabular}
    \label{spectral_paras_20201208_compmag}
\begin{list}{}{}
    \item[Note]{: Following the notes in Table~\ref{spectral_paras_20201120_compmag}.}
\end{list}
    \end{center}
\end{table*}

\begin{table*}[ptbptbptb]
    \begin{center}
\caption{Parameters of the spectral fitting on December 13 by \texttt{compmag} model}
    \begin{tabular}{cccccccccccc}
\hline
\hline
& Parameters & Average & Phase 0.0-0.1 & Phase 0.5-0.6 
\\
\hline
tbabs & $n_{\rm H}\ (10^{22}\ \rm cm^{-2})$ & $0.44$ (fixed) & $0.44$ (fixed) & $0.44$ (fixed)
\\
gabs1 & $E_{\rm cyc1}$ (keV) & $46.4_{-0.2}^{+0.1}$ & $49.1_{-0.8}^{+0.9}$ & $42.4_{-0.6}^{+0.4}$
\\
& $\sigma_{\rm cyc1}$ (keV) (line width) & $14_{-3}^{+3}$ & $14_{-2}^{+5}$ & $15_{-3}^{+4}$
\\
& $d_{\rm cyc1}$ (keV) (line deep) & $14_{-1}^{+1}$ & $22_{-3}^{+3}$ & $13_{-2}^{+1}$
\\
gabs2 & $E_{\rm cyc2}$ (keV) & $112_{-1}^{+1}$ & $112_{-10}^{+4}$ & $114_{-4}^{+1}$
\\
& $\sigma_{\rm cyc2}$ (keV) (line width) & $\geq$ 5 & $10$ (fixed) & $10$ (fixed)
\\
& $d_{\rm cyc2}$ (keV) (line deep) & $21_{-2}^{+3}$ & $12_{-8}^{+9}$ & $21_{-9}^{+1}$
\\
bbodyrad & kT (keV) & $0.46_{-0.02}^{+0.01}$ & $0.46$ (fixed) & $0.46$ (fixed)
\\
& norm & $2108_{-71}^{+199}$ & $2292_{-138}^{+76}$ & $1992_{-141}^{+175}$
\\
gaussian & $E_{\rm Fe}$ (keV) & $6.50_{-0.02}^{+0.03}$ & $6.6$ (fixed) & $6.6$ (fixed)
\\
& $\sigma_{\rm Fe}$ (keV) & $0.06_{-0.01}^{+0.01}$ & $0.06$ (fixed) & $0.06$ (fixed)
\\
& norm ($10^{-2}$)& $0.39_{-0.04}^{+0.05}$ & $0.39$ (fixed) & $0.39$ (fixed)
\\
compmag & $kT_{\rm bb}$ (keV) & $1.37_{-0.06}^{+0.03}$ & $1.37$ (fixed) & $1.37$ (fixed)
\\
(A=1, betaflag=1) & $kT_{\rm e}$ (keV) & $3.4_{-0.1}^{+0.2}$ & $3.2_{-0.1}^{+0.1}$ & $3.7_{-0.1}^{+0.1}$
\\
& $\tau$ & $0.78_{-0.05}^{+0.04}$ & $0.78$ (fixed) & $0.78$ (fixed)
\\
& $\eta$ & $0.46_{-0.01}^{+0.01}$ & $0.62$ (fixed) & $0.62$ (fixed) 
\\
& $\beta_0$ & $0.24_{-0.01}^{+0.01}$ & $0.24$ (fixed) & $0.24$ (fixed)
\\
& $r_0$ & $0.19_{-0.03}^{+0.05}$ & $0.19$ (fixed) & $0.19$ (fixed)
\\
& norm & $941_{-82}^{+115}$ & $911_{-17}^{+14}$ & $990_{-23}^{+17}$
\\
\hline
constant & ME & $0.95_{-0.01}^{+0.07}$ & $0.96_{-0.02}^{+0.01}$ & $0.97_{-0.02}^{+0.01}$
\\
constant & HE & $0.98_{-0.03}^{+0.01}$ & $0.99_{-0.02}^{+0.02}$ & $1.03_{-0.03}^{+0.02}$
\\
Fitting & $\chi^{2}$/d.o.f & 1353/1396 & 443/462 & 494/429
\\
\hline
\hline
    \end{tabular}
    \label{spectral_paras_20201213_compmag}
\begin{list}{}{}
    \item[Note]{: Following the notes in Table~\ref{spectral_paras_20201120_compmag}.}
\end{list}
    \end{center}
\end{table*}

\begin{figure*}
    \centering
    \includegraphics[width=0.9\textwidth]{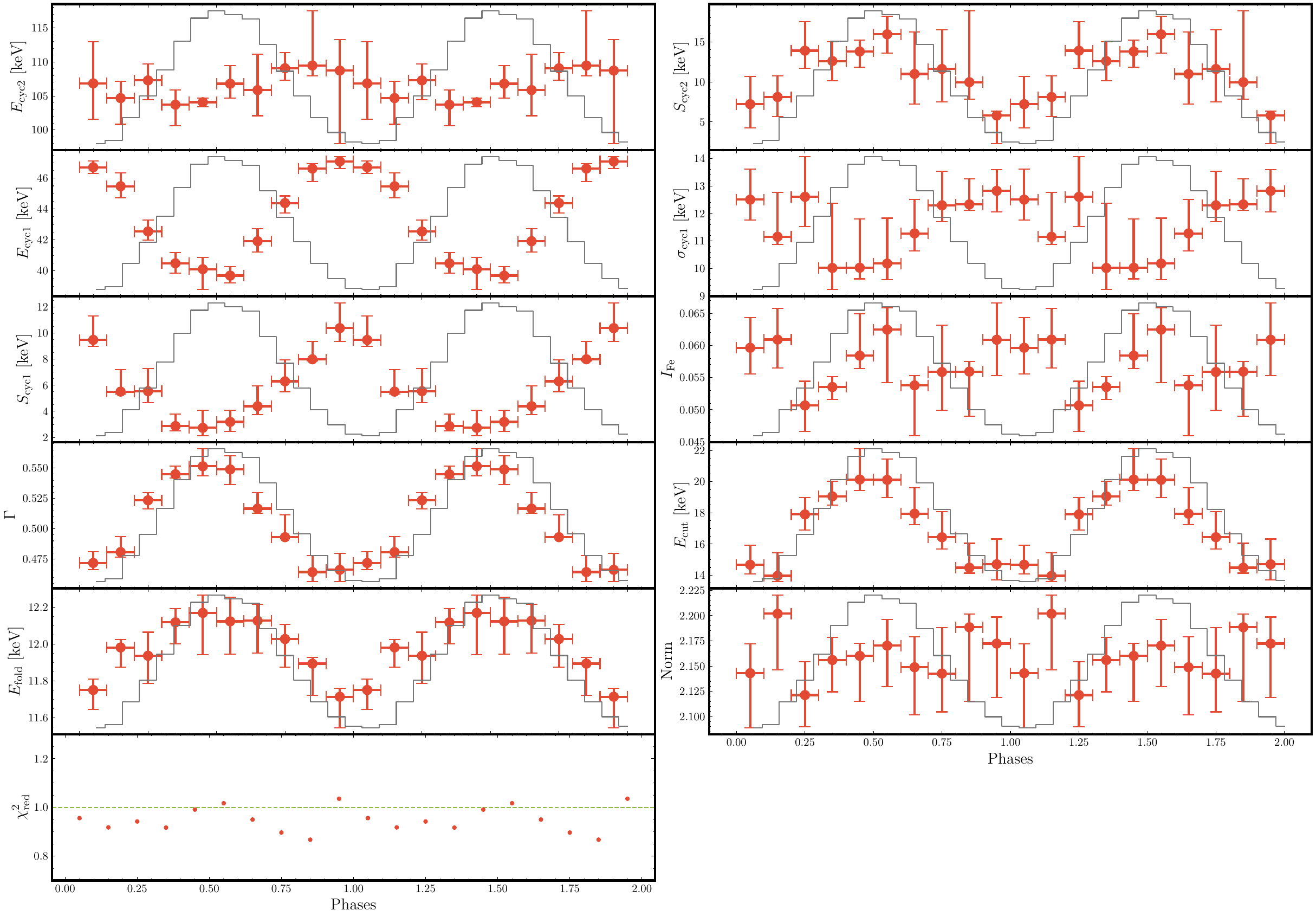}
    \caption{
    The parameters of the continuum model (\texttt{fdcut}), iron emission line (\texttt{gaussian}), and the fundamental and harmonic CRSF lines (\texttt{gabs}) from the spectral fitting results on November~20 are shown as red points. The gray lines represent the QPO profile, illustrating the correlation or lack of correlation between the QPO and the spectrum.
    }
    \label{para_20201120}
\end{figure*}

\begin{figure*}
    \centering
    \includegraphics[width=0.9\textwidth]{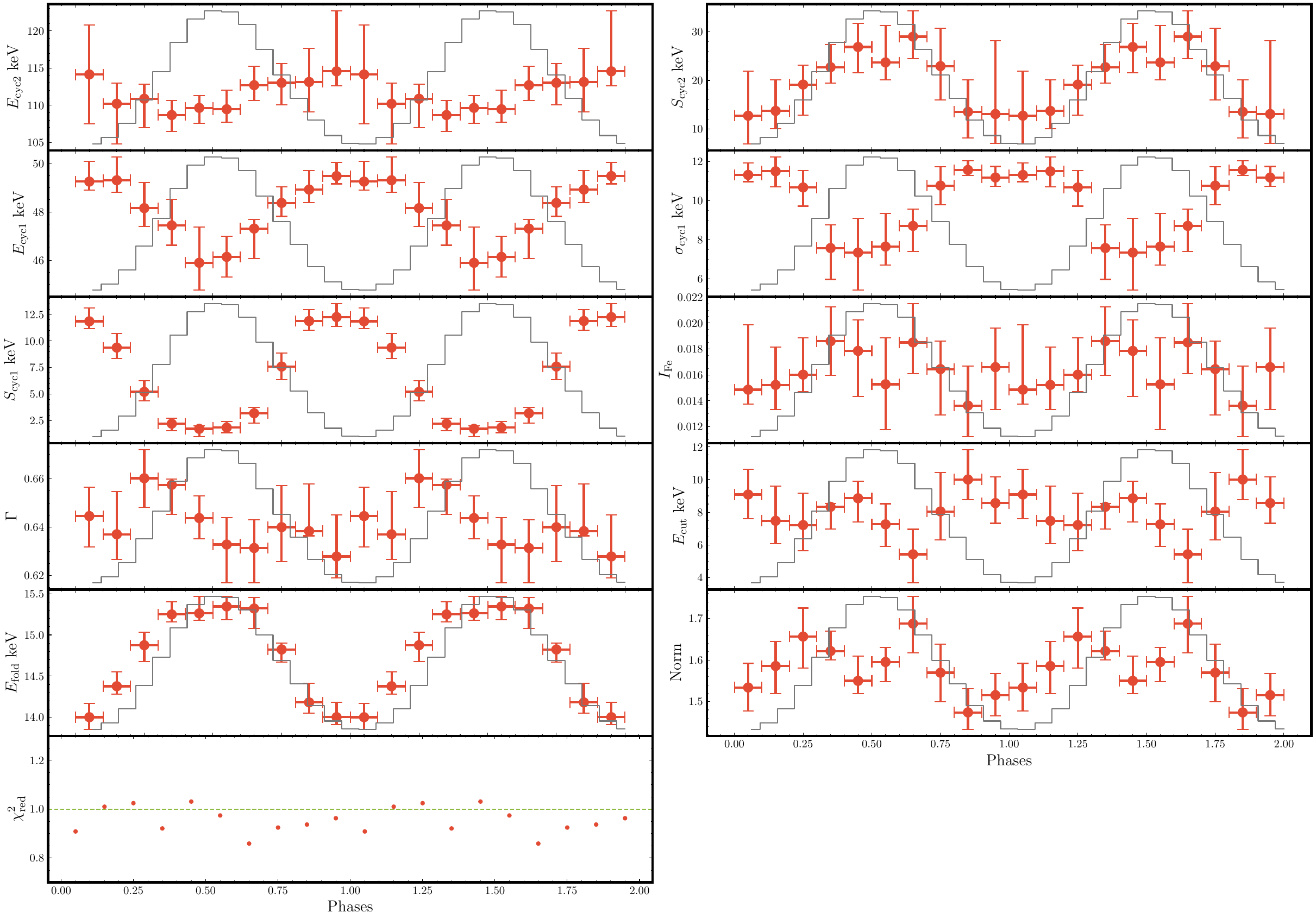}
    \caption{
    The QPO phase-resolved spectral analysis of the observation on December~8, 2020. The model is \texttt{TBabs$\times$gabs$\times$gabs$\times$(fdcut+gaussian)}.
    }
    \label{sub_para_1208}
\end{figure*}

\begin{figure*}
    \centering
    \includegraphics[width=0.9\textwidth]{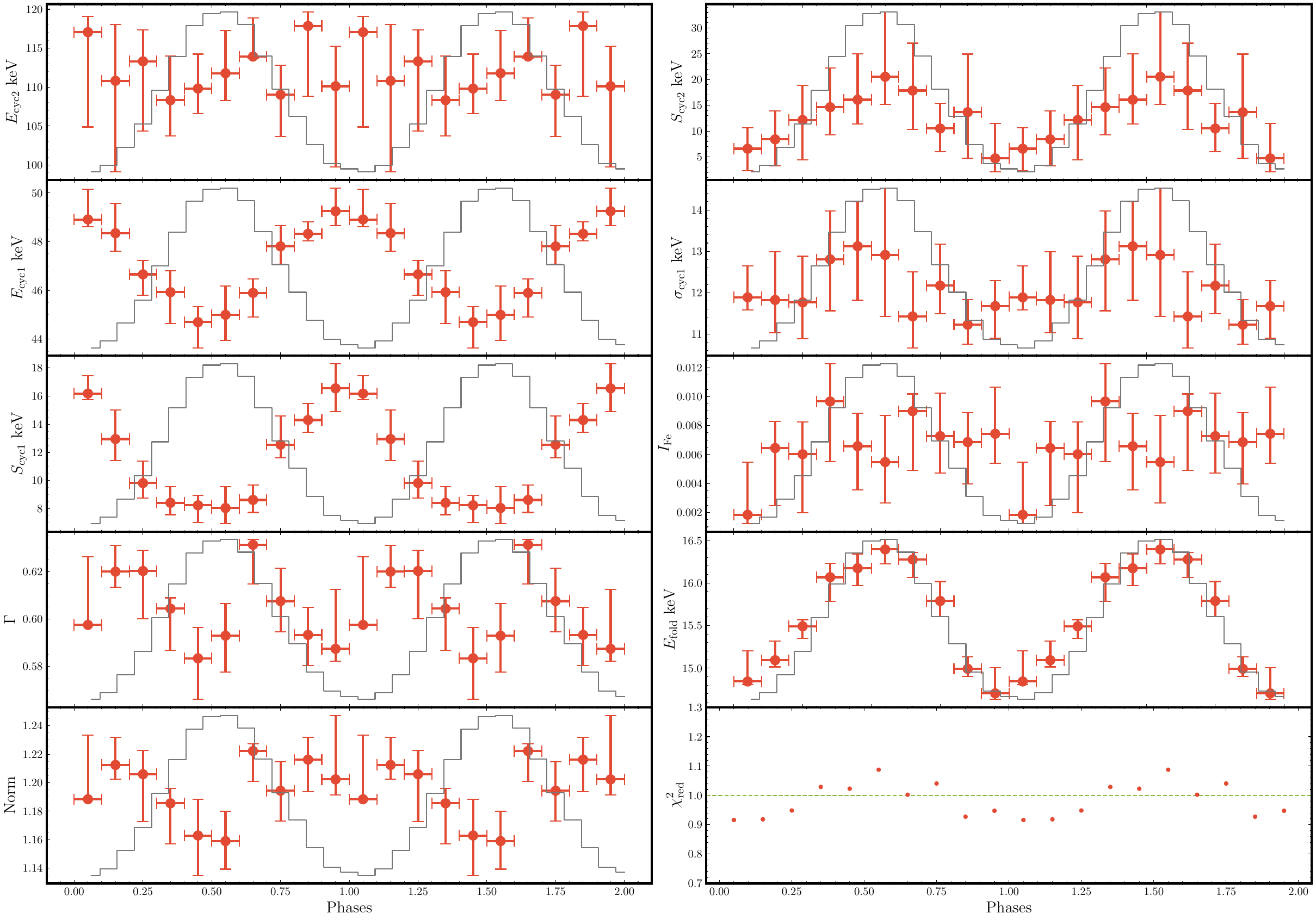}
    \caption{
    The QPO phase-resolved spectral analysis of the observation on December~13, 2020. 
    The model is \texttt{TBabs$\times$gabs$\times$gabs$\times$(fdcut+gaussian)}.
    }
    \label{sub_para_1213}
\end{figure*}

\begin{figure*}
    \centering
    \includegraphics[width=0.9\textwidth]{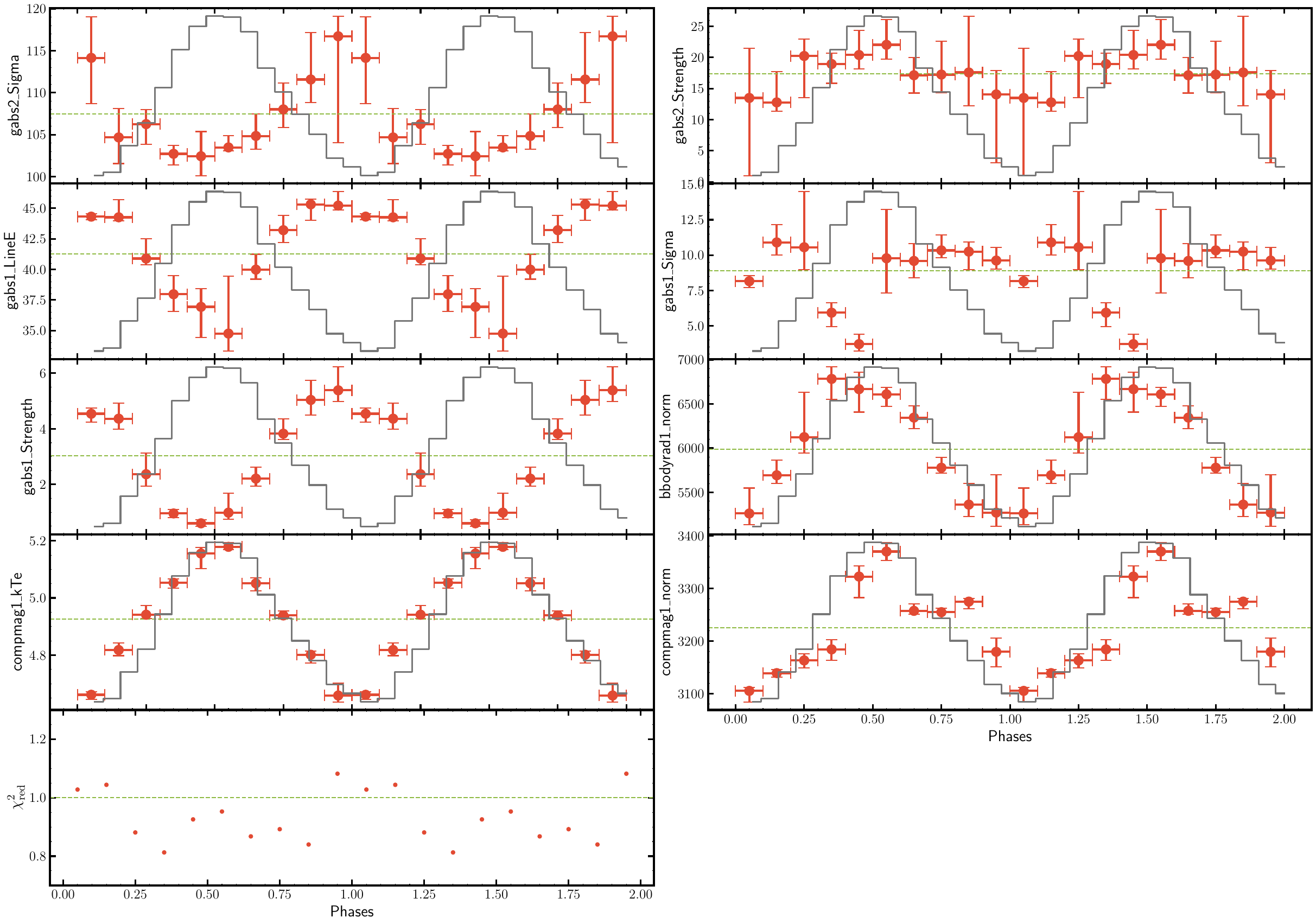}
    \caption{
    The QPO phase-resolved spectral analysis of the observation on November~20, 2020. The model is \texttt{TBabs$\times$gabs$\times$gabs$\times$(bbodyrad+gaussian+compmag)}.
    }
    \label{para_20201120_compmag}
\end{figure*}

\begin{figure*}
    \centering
    \includegraphics[width=0.9\textwidth]{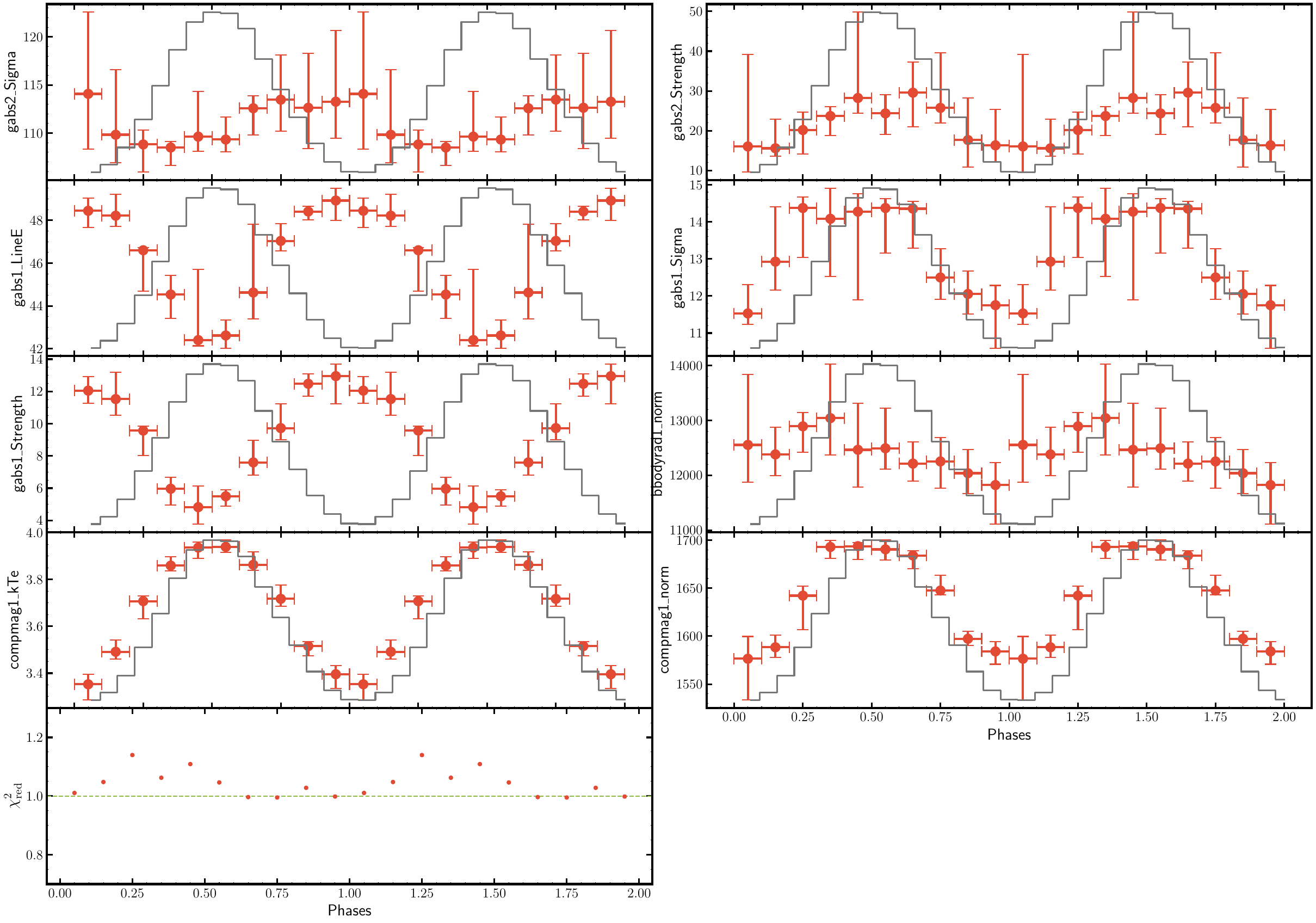}
    \caption{
    The QPO phase-resolved spectral analysis of the observation on December~8, 2020. 
    The model is \texttt{TBabs$\times$gabs$\times$gabs$\times$(bbodyrad+gaussian+compmag)}.}
    \label{sub_para_1208_compmag}
\end{figure*}

\begin{figure*}
    \centering
    \includegraphics[width=0.9\textwidth]{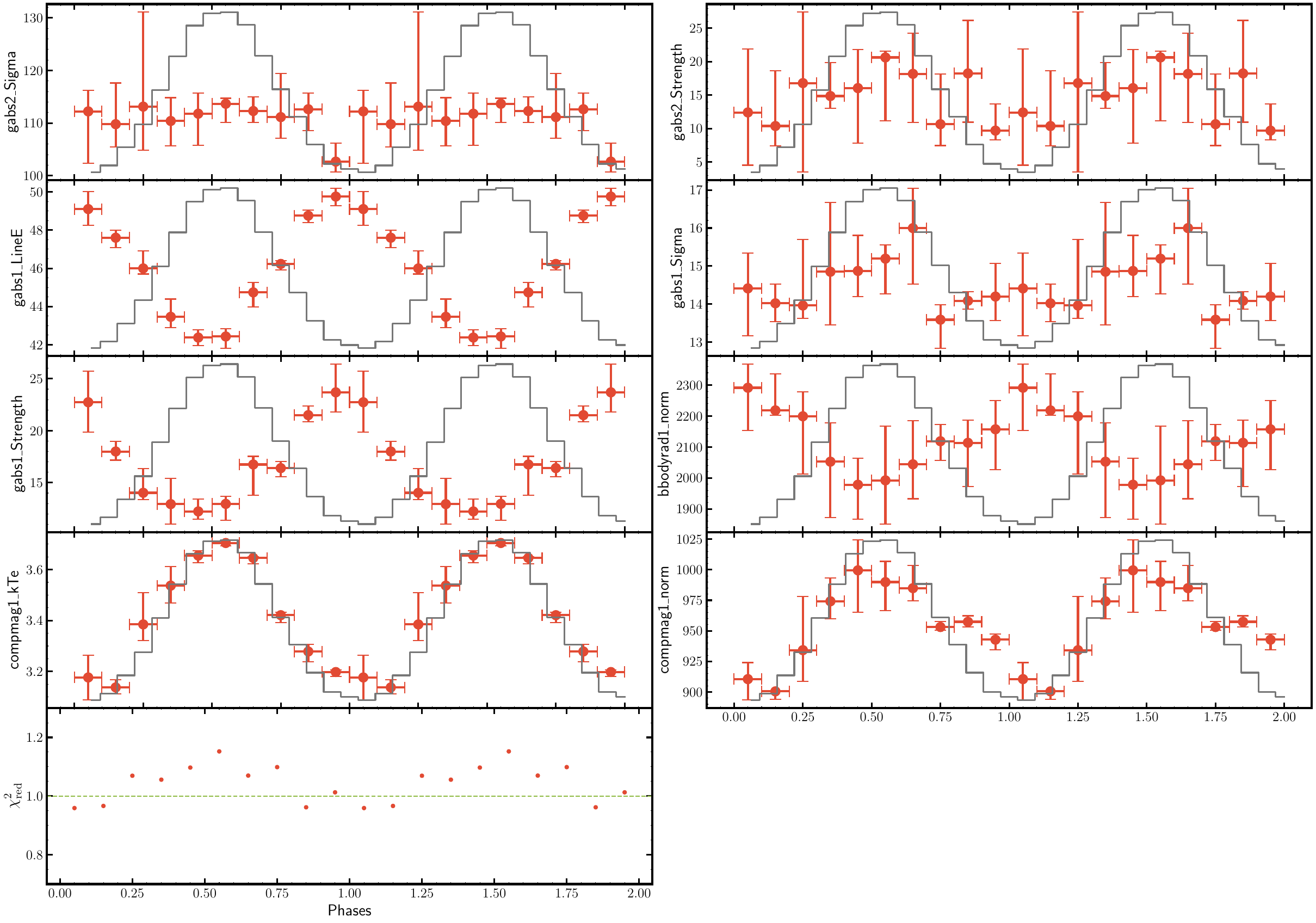}
    \caption{
    The QPO phase-resolved spectral analysis of the observation on December~13, 2020. 
    The model is \texttt{TBabs$\times$gabs$\times$gabs$\times$(bbodyrad+gaussian+compmag)}.
    }
    \label{sub_para_1213_compmag}
\end{figure*}

\begin{figure*}
    \centering
    \includegraphics[width=0.9\textwidth]{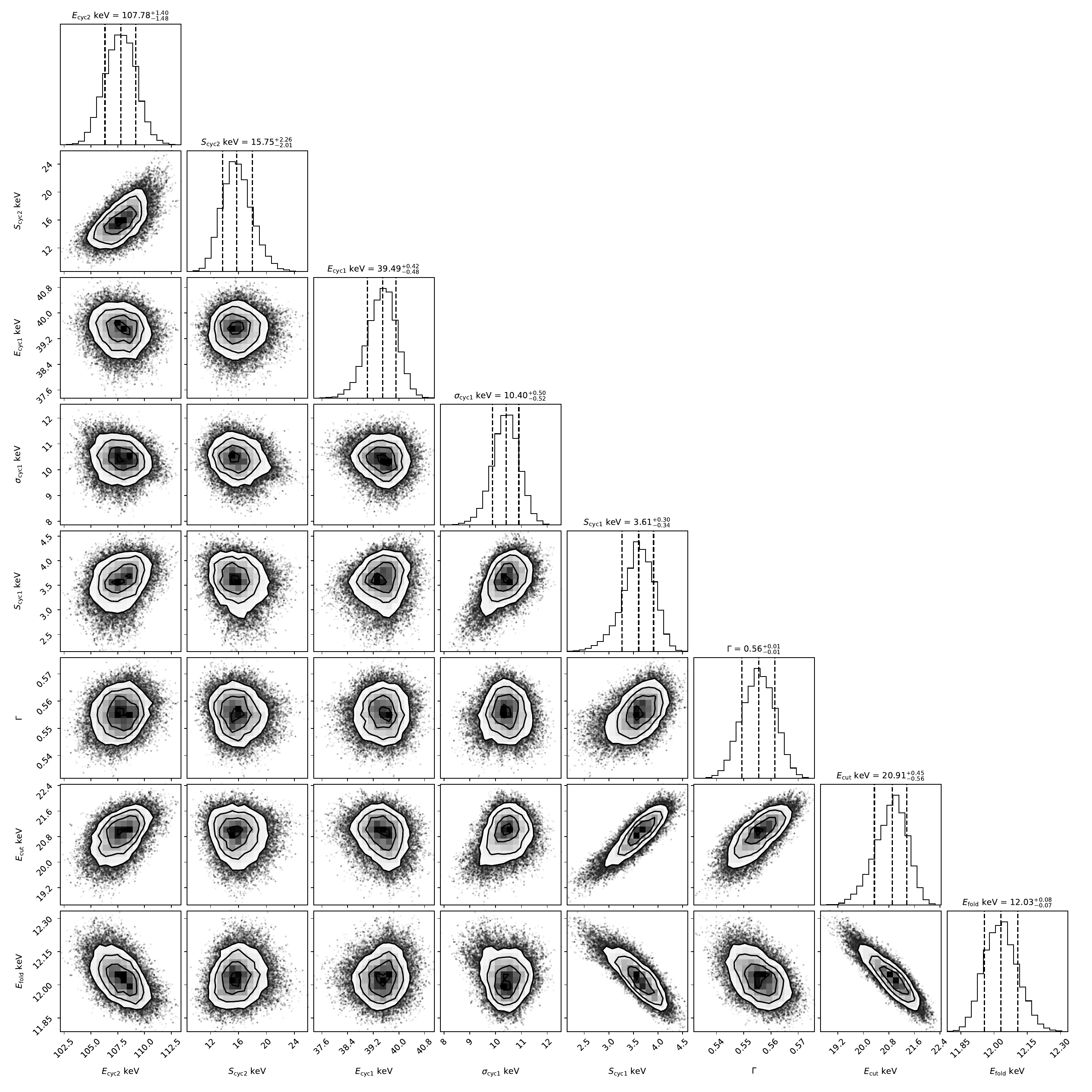}
    \caption{
    The parameter distributions from the spectral fitting of QPO phase 0.5--0.6, derived using Markov Chain Monte Carlo (MCMC) with a chain length of 100,000, are shown. Notably, the central energy ($E_{\rm cyc1}$) and width ($\sigma_{\rm cyc1}$) of the fundamental CRSF do not exhibit any significant correlations or degeneracies with the continuum model. However, the line depth shows correlations with $E_{\rm cut}$ and $E_{\rm fold}$. This indicates that the observed energy modulation of the CRSF is robust and reliable.
    }
    \label{sub_mcmc}
\end{figure*}

\begin{figure*}
    \centering
    \includegraphics[width=0.9\textwidth]{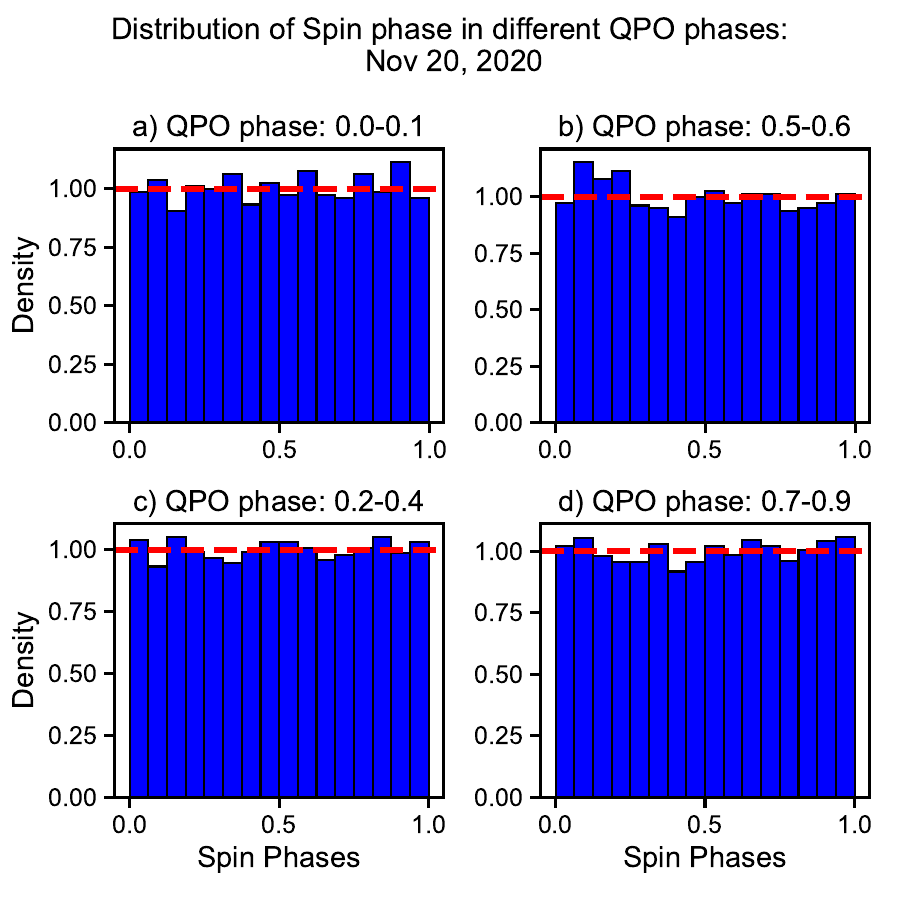}
    \caption{
    For a given QPO phase, photons from different pulse phases are uniformly distributed. This indicates that, for the QPO phase, we do not need to consider the pulse-phase dependence of the radiation.
    }
    \label{sub_pulse_distr}
\end{figure*}

\clearpage
\bibliography{sn-bibliography}

\end{document}